\documentclass[superscriptaddress,prx,twocolumn]{revtex4-1}
\usepackage{graphicx}
\usepackage{amsmath,amssymb,physics,dsfont}
\usepackage{color}
\usepackage{hyperref}
\usepackage{microtype}
\usepackage{braket}
\usepackage{defs-private}

\begin{document}

\title{Two-dimensional coherent spectrum of high-spin models via a quantum computing approach}

\author{Martin Mootz}
\email{mootz@iastate.edu}
\affiliation{Ames National Laboratory, U.S. Department of Energy, Ames, Iowa 50011, USA}

\author{Peter P. Orth}
\affiliation{Ames National Laboratory, U.S. Department of Energy, Ames, Iowa 50011, USA}
\affiliation{Department of Physics and Astronomy, Iowa State University, Ames, Iowa 50011, USA}
\affiliation{Department of Physics, Saarland University, 66123 Saarbr\"ucken, Germany}

% ...

\author{Chuankun Huang}
\affiliation{Ames National Laboratory, U.S. Department of Energy, Ames, Iowa 50011, USA}
\affiliation{Department of Physics and Astronomy, Iowa State University, Ames, Iowa 50011, USA}

\author{Liang Luo}
\affiliation{Ames National Laboratory, U.S. Department of Energy, Ames, Iowa 50011, USA}

\author{Jigang Wang}
\affiliation{Ames National Laboratory, U.S. Department of Energy, Ames, Iowa 50011, USA}
\affiliation{Department of Physics and Astronomy, Iowa State University, Ames, Iowa 50011, USA}

\author{Yong-Xin Yao}
\email{ykent@iastate.edu}
\affiliation{Ames National Laboratory, U.S. Department of Energy, Ames, Iowa 50011, USA}
\affiliation{Department of Physics and Astronomy, Iowa State University, Ames, Iowa 50011, USA}

\begin{abstract}
    We present and benchmark a quantum computing approach to calculate the two-dimensional coherent spectrum (2DCS) of high-spin models. Our approach is based on simulating their real-time dynamics in the presence of several magnetic field pulses, which are spaced in time. We utilize the adaptive variational quantum dynamics simulation (AVQDS) algorithm for the study due to its compact circuits, which enables simulations over sufficiently long times to achieve the required resolution in frequency space. 
    Specifically, we consider an antiferromagnetic quantum spin model that incorporates Dzyaloshinskii-Moriya interactions and single-ion anisotropy. 
    The obtained 2DCS spectra exhibit distinct peaks at multiples of the magnon frequency, arising from transitions between different eigenstates of the unperturbed Hamiltonian. By comparing the one-dimensional coherent spectrum with 2DCS, we demonstrate that 2DCS provides a higher resolution of the energy spectrum. 
    We further investigate how the quantum resources scale with the magnitude of the spin using two different binary encodings of the high-spin operators: the standard binary encoding and the Gray code. At low magnetic fields both encodings require comparable quantum resources, but at larger field strengths the Gray code is advantageous. Numerical simulations for spin models with increasing number of sites indicate a polynomial system-size scaling for quantum resources. Lastly, we compare the numerical 2DCS with experimental results on a rare-earth orthoferrite system. The observed strength of the magnonic high-harmonic generation signals in the 2DCS of the quantum high-spin model aligns well with the experimental data, showing significant improvement over the corresponding mean-field results.
\end{abstract}

\maketitle

\section{Introduction}

Understanding the magnetic properties of quantum materials is crucial for the development of new spintronic devices, quantum sensors, and quantum computer architectures~\cite{Li2013,Chumak2015, Jungwirth2016, Degen2017,Lingos2017,Lingos2021, deLeon2021,ang2022architectures}. Terahertz (THz) coherent spectroscopy~\cite{Kuehn2009, Kuehn2011, Junginger2012, woernerUltrafastTwodimensionalTerahertz2013, Maag2016,Yang2018TerahertzQT, Johnson2019, Higgs_2dTHz,Luo2019,Yang2019,Vaswani2020,Vaswani2020b,Yang2020lLightCS, mahmoodObservationMarginalFermi2021,  Vaswani2021,Song2023UltrafastMP} provides a powerful tool for characterizing the low-energy spin excitations of such materials, which are often inaccessible using other techniques. By measuring the material's time-dependent response to one, two or more THz pulses, which corresponds to one-dimensional coherent spectrum (1DCS), two-dimensional coherent spectrum (2DCS), and generally multi-dimensional coherent spectrum (MDCS),  the spin dynamics, energy levels, as well as interactions between the spins can be probed. 2DCS has also been discussed theoretically as a powerful probe of observing fractionalization, decoherence, and wavefunction properties in quantum spin liquids~\cite{Wan2019,Choi2020, Nandkishore2021, negahdariNonlinearResponseKitaev2023, qiang2023probing}, low-dimensional magnets~\cite{liPhotonEchoLensing2021, hartExtractingSpinonSelfenergies2023, gaoTwodimensionalCoherentSpectrum2023, simMicroscopicDetailsTwodimensional2023, liPhotonEchoFractional2023, pottsExploitingPolarizationDependence2023}, and random  magnets~\cite{parameswaranAsymptoticallyExactTheory2020}. It has been experimentally demonstrated that THz pulses enable the detection~\cite{Lu:2017} and coherent control~\cite{Kampfrath2011} of collective excitations in the spin degrees of freedom, with their quanta referred to as magnons~\cite{Pirro2021}. Simulating such experiments is crucial for interpreting experimental results and ultimately understanding the physical properties of the studied quantum materials. This is a challenging task, however, due to the large number of degrees of freedom involved and the amount of entanglement accumulated with time~\cite{SCHOLLWOCK201196,Prosen2007,Amico2008}. Here, we therefore benchmark the prospect of using quantum computers~\cite{feynman82qc}, which can naturally simulate quantum systems. 

In recent years, quantum simulations of quantum spin models have been a vibrant area of research, offering the potential to gain insights into the behavior of complex materials that are challenging to explore using classical computers~\cite{Daley2022}. Such simulations have been performed on current noisy intermediate-scale quantum (NISQ) devices~\cite{nisq, Hempel2018, hardware_efficient_vqe, Kemper_magnon, Chen2022} that are currently constrained by the number of qubits available and by the inherent noise and errors.
Quantum dynamics can be simulated using a Trotter decomposition of the time evolution operator~\cite{lloyd1996, Trotter_dynamics_Knolle}, but this method is limited to simulate early-time dynamics on NISQ devices due to the $\mathcal{O}(t^{1+1/k})$ scaling of circuit depth with time, where $k$ is the Trotter expansion order~\cite{childs2018quamtumsimulation}. Therefore, quantum circuit compression algorithms~\cite{commeau2020variational, cirstoiu2020variational, gibbs2021longtime, Peng:2022, Camps:2022, Kemper:2022, Bassman:2022, benedetti2020hardware, Berthusen:2022}, including the Variational quantum dynamics simulation (VQDS) algorithm~\cite{theory_vqs}, have been proposed to address this challenge. VQDS involves preparing a variational ansatz state that is optimized to approximate the exact time-evolved state of the system. One can derive an equation of motion governing the dynamics of the variational parameters using the McLachlan variational principle~\cite{theory_vqs, Endo20variational, nagano2023quench}, where the distance between the variational state and the exact time-evolved state is minimized. However, the accuracy of VQDS is tied to the expressibility of the variational ansatz in representing the dynamical states of the system. To address this issue, an adaptive variational quantum dynamics simulation (AVQDS) approach has been developed in Ref.~\cite{AVQDS}. To control the accuracy of AVQDS, the McLachlan distance is kept below a threshold along the dynamical path by adaptively appending new parametrized unitaries to the variational ansatz with generators selected from a predefined operator pool. AVQDS has been demonstrated to generate accurate quantum dynamics simulation results with greatly compressed circuits  measured by the number of entangling gates compared to Trotterized state evolution~\cite{AVQDS}.

\begin{figure}[t!]
\begin{center}
		\includegraphics[scale=0.60]{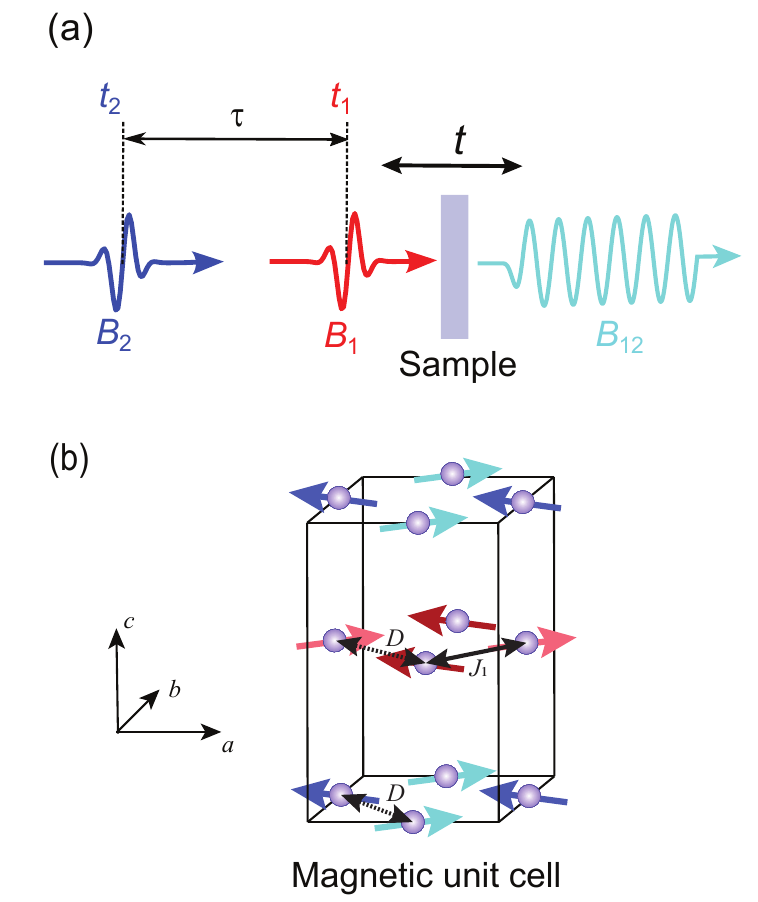}
		\caption{\textbf{Two-dimensional coherent spectroscopy of high-spin systems.} (a) Schematic representation of the two-dimensional spectroscopy configuration considered in this paper. The sample is excited by two collinear phase-locked few-cycle magnetic field pulses. The magnetic field pulse denoted by $B_{1}$ (red) is centered at $t_{1}=0$ while pulse $B_{2}$ (blue) is centered at $t_\mathrm{2}=\tau$ with inter-pulse delay $\tau=t_{2}-t_{1}$. $B_{12}$ (cyan) is the transmitted magnetic field after excitation with both pulses. (b) High spin model for rare-earth (R) orthoferrites RFeO$_3$ considered in the simulations. The positions of the iron ions in the magnetic unit cell are indicated by balls while the corresponding spins are shown as arrows. The orthorhombic crystal consists of four iron sublattices indicated by the different colored arrows with approximately antiferromagnetic order between nearest-neighbor spins within the $ab$-plane and ferromagnetic order along $c$-axis. Examples for nearest neighbor antiferromagnetic interaction and DM interaction are indicated by solid and dotted double arrows, respectively. Since the spin dynamics is dominated by the iron sublattice, the positions of R and O atoms are not shown.} 
		\label{fig1} 
\end{center}
\end{figure}

In this work, we use AVQDS to calculate the 2DCS of high spin-models. To model 2DCS experiments, we consider the phase-locked collinear 2-pulse geometry illustrated in Fig.\figref{fig1}{(a)}, where the sample is excited by two equal few-cycle magnetic field pulses separated by the inter-pulse delay $\tau$. Compared to Ref.~\cite{AVQDS}, we here utilize a high-order integrator to calculate the dynamics of the variational parameters. This results in shallower quantum circuits and fewer time steps. We compare two different binary encodings of the high-spin operators, the standard binary encoding and the Gray code, and determine the scaling of the required quantum resources with spin magnitude $s$. We further analyze how the quantum resources scale with the number of sites for quantum spin models of spin magnitude $s\in \{1/2, 1, 3/2\}$.

We consider an antiferromagnetic quantum high-spin model, including Dzyaloshinskii-Moriya (DM) interaction and single-ion anisotropy. This model has been previously used to simulate the magnetic properties of rare-earth orthorferrites, based on a classical description of the spins~\cite{Herrmann:1964, Shane:1968, Hahn:2014, Lu:2017}. 
The orthorhombic magnetic unit cell of these materials shown in Fig.\figref{fig1}{(b)} is characterized by a nearly antiferromagnetic arrangement of nearest-neighbor spins within the $ab$-plane with spin $s\approx 5/2$
and weak ferromagnetic order along the $c$-axis, induced by the DM interaction along the $b$-axis. Using a simplified two-site model, we calculate 2DCS spectra for spin $s=1$ and demonstrate that 2DCS provides higher resolution of the energetics of the high-spin model compared to conventional 1DCS. Through the application of a susceptibility expansion, we show that the peaks in two-dimensional (2D) frequency space arise from transitions between different eigenstates of the unperturbed Hamiltonian. Finally, we compute the 2DCS spectrum for a two-site spin-$s=5/2$ model to compare it with the results of a 2DCS experiment performed on a rare-earth orthoferrite system. The observed strength of the magnonic high-harmonic generation signals in the simulated 2DCS spectra of the quantum high-spin model agrees well with the experiment, in contrast to simulations based on the corresponding mean-field model.

The paper is organized as follows. In section~\ref{sec:model}, we introduce the high-spin model and the simulation setup. We present the two binary encodings of high spin-$s$ operators considered in this paper in section~\ref{sec:S_trans}. Section~\ref{sec:AVQDS} provides a summary of the key concepts of the AVQDS approach with a high-order Runge-Kutta integrator. In section~\ref{sec:1Dvs2D}, we compare the results of 1DCS and 2DCS, and benchmark the performance of AVQDS. The analysis of quantum resource scaling with increasing spin $s$ and number of sites are presented in sections~\ref{sec:S-dep} and \ref{sec:N-dep}, respectively. Finally, in section~\ref{sec:4-site}, we compare the 2DCS simulation results of a two-site spin-$s=5/2$ model with experimental 2DCS results. The paper concludes with a summary of findings and an outlook.

\section{Model and Simulation setup \label{sec:model}}

To model 2DCS spectroscopy experiments on high-spin materials and to benchmark the performance of the AVQDS method with a high-order integrator, we study a high-spin model which  is used to analyze the THz light-driven spin dynamics in rare-earth orthoferrites~\cite{Lu:2017}. The magnetic properties of these materials are characterized by the orientation of the magnetic moment on the iron sites which have spin $s=5/2$. The orthorhombic crystal consists of four iron sublattices with dominant antiferromagnetic order along $a$-axis as illustrated in Fig.~\figref{fig1}{b}. The spins of the antiferromagnetic order are slightly canted along the crystal $c$-axis which leads to a net magnetization and thus ferromagnetic order along the $c$-axis. This spin canting is introduced by the Dzyaloshinskii-Moriya (DM) interaction in most rare-earth orthoferrites, while it can also be induced by strong single-ion anisotropy of the orthorhombic crystal~\cite{Herrmann:1964}. In this paper, we focus on ferromagnetic order predominantly induced by the DM interaction. In this situation, the low-energy magnetic response can be accurately described by a two-sublattice model~\cite{Herrmann:1964}, which we consider in this paper. Compared to previous works~\cite{Herrmann:1964,Shane:1968,Hahn:2014,Lu:2017} where the magnetic properties and light-driven spin dynamics of rare-earth orthoferrites are described based on a fully classical treatment of the spins, we study the magnetic field driven spin dynamics of the corresponding quantum spin model with $N$ sites and open boundary conditions which reads:
\begin{align}
\label{eq:Ham}
\hat{\mathcal{H}} =&\,J\sum_{i=0}^{N-2}\hat{\mathbf{S}}_i\cdot\hat{\mathbf{S}}_{i+1}-\mathbf{D}\cdot\sum_{i=0}^{N-2}\hat{\mathbf{S}}_i\times\hat{\mathbf{S}}_{i+1} \nonumber \\
&-K_a\sum_{i=0}^{N-1}(\hat{S}^x_i)^2-K_c\sum_{i=0}^{N-1}(\hat{S}^z_i)^2-\mathbf{B}\cdot\sum_{i=0}^{N-1}\hat{\mathbf{S}}_i\,.
\end{align}
Here, $\hat{S}^j_i$ ($j=x,y,z$) are spin operators at site $i$ of generic spin $s$. Note that we identify $x=a$, $y=b$, and $z=c$.  The first term of the Hamiltonian represents the antiferromagnetic coupling between nearest-neighbor (solid double arrow in Fig.\figref{fig1}{(b)}) with exchange constant $J > 0$. The second term of equation~\eqref{eq:Ham} accounts for the DM interaction (dotted double arrow in Fig.~\figref{fig1}{(b)}) with anti-symmetric exchange vector $\mathbf{D}$ which we assume to be aligned along $b$-direction, i.~e., $\mathbf{D}=D\,\hat{\mathbf{y}}$. The single-ion anisotropy of the orthorhombic crystal is modelled by the third and fourth terms of the Hamiltonian, where anisotropy constants $K_a$ and $K_c$ are associated with the $a$- and $c$-directions, respectively. The last term of the Hamiltonian~\eqref{eq:Ham} describes the Zeemann interaction with a spatially uniform magnetic field $\mathbf{B}(t)$, which is chosen to be polarized along the $z$-axis.

We model the transmitted magnetic field measured in 2DCS experiments~\cite{Lu:2017} by calculating the time evolution of the magnetization $\mathbf{M}(t)\equiv \langle \hat{\mathbf{S}}^\mathrm{tot}\rangle=\langle\Psi[t]|\hat{\mathbf{S}}^\mathrm{tot}|\Psi[t]\rangle$, where $\hat{\mathbf{S}}^\mathrm{tot}=\sum_j \hat{\mathbf{S}}_j$ is the total spin operator and $|\Psi[t]\rangle$ is the quantum state of system at time $t$. Since the magnetic field induces a net magnetization along $c$-direction, while the total magnetization along $a$ and $b$-directions are zero for magnetic fields polarized along $c$-direction, we focus on the dynamics of $M^z(t)$ in this paper. To simulate 2DCS experiments on the high-spin model~\eqref{eq:Ham}, we consider a collinear geometry where the magnetic field consists of two copropagating pulses that are separated in time by $\tau$: $\mathbf{B}(t) = \mathbf{B}_1 f(t) + \mathbf{B}_2 f(t - \tau)$, where $f(t)$ is the narrow lineshape of the few-cycle pulse (Fig.\figref{fig1}{a}). We compute the nonlinear differential magnetization ${M}^z_\mathrm{NL}$, which corresponds to the transmitted magnetic field $B^z_\mathrm{NL}$ in 2DCS experiments~\cite{Lu:2017}. This is done by determining the magnetization induced by both pulses, $M^z_{12}(t,\tau)$, as a function of time $t$ and the inter-pulse delay $\tau$, as well as the magnetizations resulting from only applying pulse 1 (red line in Fig.\figref{fig1}{a}), ${M}^z_{1}(t)$, and from only applying pulse 2 (blue line in Fig.\figref{fig1}{(a)}), ${M}^z_{2}(t,\tau)$. The nonlinear differential magnetization is then given by
\begin{align}
\label{eq:Mnl}
M^z_\mathrm{NL}(t,\tau)=M^z_\mathrm{12}(t,\tau)-M^z_\mathrm{1}(t)-M^z_\mathrm{2}(t,\tau)\,.    
\end{align} 
Here we note that the nonlinear differential magnetization~\eqref{eq:Mnl} cannot be solved analytically exactly for the studied high-spin models, requiring the use of numerical methods such as exact diagonalization or AVQDS for its calculation. To obtain the 2DCS spectra, we perform a Fourier transform of $M^z_\mathrm{NL}(t,\tau)$ with respect to $t$ (frequency $\omega_t$) and $\tau$ (frequency $\omega_\tau$). The 2D spectrum is characterized by the intensity at frequency coordinate $(\omega_t,\omega_\tau)$.

\section{Qubit representation of spin-$s$ model \label{sec:S_trans}}
To simulate the driven spin system using quantum computer, the high-spin operators of Hamiltonian~\eqref{eq:Ham} need to be converted to multi-qubit operators. Therefore, we adopt the scheme of~\cite{sawaya2020} for implementing a generic encoding. The spin operator $\hat{S}$ of spin $s$ can be written as
\begin{align}
    \hat{S}^j=\sum_{l,l'=0}^{d-1}a^j_{l,l'}|l\rangle\langle l'|\,,
\label{eq:Sj}
\end{align}
with $j=\{x,y,z\}$. Here $d=2s+1$ represents the number 
of elements in the basis set of the quantum spin of magnitude $s$, while $|l\rangle$, $l=0,\dots,d-1$, denote the basis states of the spin. The coefficients $a^j_{l,l'}$ for the different spin components are defined by
\begin{align}
    a^x_{l,l'}\equiv &\langle l|\hat{S}^x|l'\rangle=\frac{1}{2}\left(\delta_{l,l'+1}+\delta_{l+1,l'}\right)\nonumber \\
    & \qquad\qquad\times\sqrt{(s+1)(l+l'+1)-(l+1)(l'+1)}\,, \nonumber \\
    a^y_{l,l'}\equiv &\langle l|\hat{S}^y|l'\rangle=\frac{i}{2}\left(\delta_{l,l'+1}-\delta_{l+1,l'}\right)\nonumber \\
    & \qquad\qquad\times\sqrt{(s+1)(l+l'+1)-(l+1)(l'+1)}\,, \nonumber \\
    a^z_{l,l'}\equiv &\langle l|\hat{S}^z|l'\rangle=(s-l)\delta_{l,l'}\,.
\end{align}
To encode the integer $l$ of quantum state $|l\rangle$ in equation~(\ref{eq:Sj}), a binary encoding $\mathcal{R}(l)$ is introduced which converts $l$ to a bit string on $N_\mathrm{q}$ bits $x_{N_\mathrm{q}-1}\dots x_1 x_0$ where $x_i\in\{0,1\}$. As a result, the state $|l\rangle$ can be encoded via 
\begin{align}
    |l\rangle 	\longmapsto |\mathcal{R}(l)=|x_{N_\mathrm{q}-1}\rangle\dots |x_0\rangle\,.
\end{align}
The transformation of the operator $|l\rangle\langle l'|$ to qubit space is then given by
\begin{align}
    |l\rangle\langle l'| \longmapsto
    \bigotimes_{j=0}^{N_\mathrm{q}-1}|x_j\rangle\langle x_j'|_j\,.
\end{align}
The operators $|x\rangle\langle x'|$ are transformed to qubit operators by using (for every qubit $j$)
\begin{align}
\label{eq:qubit_ops}
    &|0\rangle\langle 1|=\frac{1}{2}(\sigma^x+i\sigma^y)\equiv \sigma^{-}\,, \quad |1\rangle\langle 0|=\frac{1}{2}(\sigma^x-i\sigma^y)\equiv \sigma^{+}\,, \nonumber \\
    &|0\rangle\langle 0|=\frac{1}{2}(I+\sigma^z)\,,\quad |1\rangle\langle 1|=\frac{1}{2}(I-\sigma^z)\,,
\end{align}
where $I$ is the identity and $\{\sigma^a\}$ are Pauli matrices.

In this paper, we consider two different encodings $\mathcal{R}(l)$: (i) The standard binary encoding (std), which uses base-two numbering such that the integer $l$ is expressed as $l\mapsto x_0 2^0+x_1 2^1+x_2 2^2\dots$. (ii) The Gray code~\cite{Siwach:2021,matteo2021}, where the Hamming distance, measured by the number of bits different in two bit strings, is always 1 between adjacent integer bit strings. As a result, only one spin flip is needed to transform between neighboring states.
Both encodings, standard binary encoding and Gray code, use the full Hilbert space of the qubit basis and require $N_\mathrm{q}=\lceil\log_2 d\rceil$ qubits for a $d$-level operator where $\lceil\; \cdot \; \rceil$ is the ceiling function. An example of the spin-$s$ encodings is shown in Table~1 for $s=5/2$, which has $d=6$ levels and the mapping requires 3 qubits. The spin-$s$ operators~(\ref{eq:Sj}) expressed in terms of the qubit operators~\eqref{eq:qubit_ops} are explicitly given in appendix~\ref{sec:S_trafo} for $s=1,\,3/2,\,2,\,$ and 5/2. 

It is important to note that the mapping from the basis states of the spin-$s$ system to qubit basis states is in general not bijective. As a result, the ground state preparation of high-spin models with quantum computing approaches such as qubit-ADAPT VQE~\cite{MayhallQubitAVQE} and AVQITE~\cite{AVQITE} might break spin conservation in some applications due to the redundant subspace in the qubit Hilbert space. However, the ground states prepared using qubit-ADAPT VQE and AVQITE in this paper have been carefully checked and are always in the correct spin-$s$ subspace for all studied high-spin models. Specifically, we calculated the expectation value of the local spin operator $S_j^2=(S^x_j)^2+(S^y_j)^2+(S^z_j)^2$, which for the exact ground state is given by $\langle S_j^2\rangle=s(s+1)$. The relative error of $\langle S_j^2\rangle$ for the prepared ground state is smaller than $2.7\times 10^{-5}$ for all the studied high spin models. This demonstrates the conservation of the spin and confirms that the obtained ground states prepared using qubit-ADAPT VQE and AVQITE lie within the correct spin-$s$ sectors. This is consistent with the infidelity of the ground state preparation discussed in sections~\ref{sec:1Dvs2D}, \ref{sec:S-dep}, and \ref{sec:4-site}, which is smaller than $6.5\times 10^{-3}$ for all studied high spin models. One can understand why the redundant subspace due to the non-bijective encodings has negligible effect on the ground state preparation by examining the spectrum of the encoded Hamiltonian in qubit space. Since the Hamiltonian does not act on the basis states of the redundant subspace, they are decoupled from the spin-$s$ subspace and merely contribute to eigenstates at energy zero. Because for all our studied models, the ground state energy is much lower ($E_0/J < -2.0$), the ground state preparation using qubit-ADAPT VQE or AVQITE is hardly affected by the redundant subspace. This is also consistent with the results presented in Ref.~\cite{getelina2024}. 

\begin{table}[]
    \centering
    \begin{tabular}{||c|c |c||} 
		\hline
		Spin level & Standard binary & Gray code \\ [0.5ex] 
		\hline\hline
		$|0\rangle$ & $|000\rangle$ & $|000\rangle$ \\ 
		\hline
		$|1\rangle$ & $|001\rangle$ & $|001\rangle$ \\ 		
		\hline
		$|2\rangle$ & $|010\rangle$ & $|011\rangle$ \\ 
		\hline
		$|3\rangle$ & $|011\rangle$ & $|010\rangle$ \\  
		\hline
		$|4\rangle$ & $|100\rangle$ & $|110\rangle$ \\ 
		\hline
		$|5\rangle$ & $|101\rangle$ & $|111\rangle$ \\ 
		\hline
	\end{tabular}
    \caption{Standard binary encoding and Gray code for spin $s=5/2$.}
    \label{tab:4wm}
\end{table}
	

\section{AVQDS with high-order integrator \label{sec:AVQDS}}
\subsection{AVQDS algorithm}
The quantum computing approach used in this paper, AVQDS, was introduced in Ref.~\cite{AVQDS}, and we here provide a summary of its key concepts. 
For a system initially in a pure state $|\Psi\rangle$ and described by a time-dependent Hamiltonian $\hat{\mathcal{H}}$, the dynamics of the density matrix $\rho=|\Psi\rangle\langle\Psi|$ is governed by the von Neumann equation
\begin{align}
	\frac{\mathrm{d}\rho}{\mathrm{d}t}=\mathcal{L}[\rho]\,,
\end{align}
where $\mathcal{L}[\rho]=-i\left[\hat{\mathcal{H}},\rho\right]$. The state $|\Psi\rangle$ is represented in a parameterized form $|\Psi[\bth]\rangle$ with a real-valued time-dependent variational parameter vector $\bth(t)$ of dimension $N_\theta$~\cite{theory_vqs}. The dynamics of $\bth$ follows from the equations of motion determined by the McLachlan variational principle~\cite{mclachlan64variational}. It minimizes the squared McLachlan distance $\mathcal{L}^2$, which is defined as the Frobenius norm of the difference between the exact and the variational state propagation: 
\begin{align}
\label{eq:L2}
    \mathcal{L}^2&\equiv\bigg\|\sum_\mu\frac{\partial\rho[\bth]}{\partial\theta_\mu}\dot{\theta}_\mu-\mathcal{L}[\rho]\bigg\|^2_\mathrm{F}\nonumber \\
    &=\sum_{\mu\nu}M_{\mu\nu}\dot{\theta}_\mu\dot{\theta}_\nu-2\sum_\mu V_\mu\dot{\theta}_\mu+\mathrm{Tr}[\mathcal{L}^2(\rho)]\,.
\end{align}
The elements of the $N_\theta\times N_\theta$ matrix $M$ and vector $V$ of dimension $N_\theta$ are defined by
\begin{align}
\label{eq:M}
M_{\mu,\nu}\equiv &\mathrm{Tr}\left[\frac{\partial]\rho[\bth]}{\partial\theta_\mu}\frac{\partial]\rho[\bth]}{\partial\theta_\nu}\right]=2\mathrm{Re}\left[\frac{\partial\langle\Psi[\bth]|}{\partial \theta_\mu}\frac{\partial|\Psi[\bth]\rangle}{\partial \theta_\nu}\right.\nonumber \\ &\left.+\frac{\partial\langle\Psi[\bth]|}{\partial \theta_\mu}|\Psi[\bth]\rangle\frac{\partial\langle\Psi[\bth]|}{\partial \theta_\nu}|\Psi[\bth]\rangle\right]\,, \\
	V_\mu=&2\mathrm{Im}\left[\frac{\partial\langle\Psi[\bth]|}{\partial \theta_\mu}H|\Psi[\bth]\rangle+\langle\Psi[\bth]|\frac{\partial|\Psi[\bth]\rangle}{\partial \theta_\mu}\langle H\rangle_{\bth}\right]\,,
 \label{eq:V}
\end{align}
with $\langle H\rangle_{\bth}=\langle\Psi[\bth]|H|\Psi[\bth]\rangle$. The real symmetric matrix $M$ has a one-to-one correspondence to the quantum Fisher information matrix~\cite{meyer2021fisher}, with the second term within the bracket resulting from the global phase contribution~\cite{theory_vqs}. By expressing the last term of equation~(\ref{eq:L2}) in terms of the variance of $\hat{\mathcal{H}}$ in the variational state $|\Psi[\bth]\rangle$,
\begin{align}
    \mathrm{Tr}[\mathcal{L}^2(\rho)]=2\left(\langle\hat{\mathcal{H}}^2\rangle_{\bth}-\langle\hat{\mathcal{H}}\rangle^2_{\bth}\right)=2\,\mathrm{var}_{\bth}[\hat{\mathcal{H}}]\,,
\end{align}   
the minimization of $\mathcal{L}^2$ with respect to $\{\dot{\theta}_\mu\}$ yields
\begin{align}
\label{eq:theta_eom}
    \sum_\nu M_{\mu\nu}\dot{\theta}_\nu=V_\mu\,.
\end{align}
The above equation of motion determines the dynamics of the variational parameters whose numerical integration will be discussed in more detail below.  Based on equation~(\ref{eq:theta_eom}), the optimized McLachlan distance $\mathcal{L}^2$ of the variational ansatz $|\Psi[\bth]\rangle$ is given by
\begin{align}
    \mathcal{L}^2=2\,\mathrm{var}_{\bth}[\hat{\mathcal{H}}]-\sum_{\mu\nu}V_\mu M^{-1}_{\mu\nu}V_\nu\,,
\end{align}
which measures the accuracy of the variational dynamics.

In AVQDS, the ansatz takes a pseudo-Trotter form:
\begin{align}
\label{eq:ansatz}
|\Psi[\bth]\rangle=\prod_{\mu=0}^{N_\theta-1}e^{-i\theta_\mu\hat{\mathcal{A}}_\mu}|\Phi_0\rangle\,,  
\end{align}
with Hermitian generators $\hat{\mathcal{A}}_\mu$ ($\mu=0,\cdots,N_\theta-1)$ and reference state 
$\ket{\Phi_0}$. To ensure that the McLachlan distance $\mathcal{L}^2$ remains below a certain threshold $\mathcal{L}^2_\mathrm{cut}$ during the time evolution, additional operators from a predefined operator pool will be dynamically appended when necessary, thus expanding the set of generators in equation~\eqref{eq:ansatz}. Specifically, the McLachlan distance $\mathcal{L}^2$ is calculated for a new variational ansatz of the form $e^{-i\theta_\nu\hat{\mathcal{A}_\nu}}|\Psi[\bth]\rangle$ where the generator $\hat{\mathcal{A}}_\nu$ is chosen from the predefined operator pool of size $N_\mathrm{p}$. It is important to note that $\theta_\nu$ is initially set to zero at the current time step, which guarantees that the wave function remains smooth during time evolution. The resulting new ansatz with $\theta_\nu=0$ does not change the state of the ansatz, but can modify the McLachlan distance $\mathcal{L}^2$ since the derivative with respect to $\theta_\nu$ is non-zero. The McLachlan distance $\mathcal{L}^2_\mu$ is calculated for all $\hat{\mathcal{A}}_\mu$ of the predefined operator pool and the $\hat{\mathcal{A}}_\nu$ which produces the smallest $\mathcal{L}^2_\nu$ is selected, increasing $N_\theta \rightarrow N_\theta + 1$ by one. The ansatz adaptive procedure is continued until the McLachlan distance $\mathcal{L}^2$ of the new ansatz is smaller than $\mathcal{L}^2_\mathrm{cut}$. A threshold value of $\mathcal{L}^2_\mathrm{cut} = 10^{-5}$ is found to be sufficient in achieving accurate 2DCS simulation results in this work.

\subsection{High-order integrator}
\label{subsec:high_order_integrator}
The variational parameters are then evolved in time by inverting equation~\eqref{eq:theta_eom} which leads to
\begin{align}
	\dot\bth=M^{-1}V \equiv f(\bth[t],t)\,.
	\label{eq:ode}    
\end{align}
Solving equation~(\ref{eq:ode}) involves the numerical integration of a system of ordinary differential equations. In contrast to~\cite{AVQDS} where the Euler method was used, here we employ a fourth-order Runge-Kutta method which yields
\begin{align}
\label{eq:rk4}
	\bth[t+\delta t]=\bth[t]+\frac{1}{6}\left(k_1+2k_2+2k_3+k_4\right)\delta t\,,
\end{align}
with
\begin{align}
\label{eq:rk4-2}
	&k_1=f\left(\bth[t],t\right)\,, \nonumber \\
	&k_2=f\left(\bth[t]+\frac{k_1}{2}\delta t,t+\frac{1}{2}\delta t\right)\,, \nonumber \\
	&k_3=f\left(\bth[t]+\frac{k_2}{2}\delta t,t+\frac{1}{2}\delta t\right)\,, \nonumber \\
	&k_4=f\left(\bth[t]+k_3 \delta t,t+\delta t\right)\,.
\end{align}
The  truncation error at a single time step is of order $(\delta t)^4$. To demonstrate the advantage of this Runge-Kutta method over lower-order integration methods in AVQDS simulations, we compare to the Euler method, where 
\begin{align}
	\bth[t+\delta t]=\bth[t]+f(\bth[t],t)\,\delta t\,.
\end{align}
The truncation error of this method at a single time step is of order $(\delta t)^2$. Even though the fourth-order Runge-Kutta method introduces an computational overhead as each step involves four micro-steps to update the variational parameters compared to a single step in Euler's method, the total number of time steps and the circuit complexity can still be much reduced as demonstrated in appendix~\ref{sec:Euler}. Importantly, the time step $\delta t$  in the AVQDS simulations is dynamically adjusted such that $\max_{0\leq\mu<N_\theta}|\delta\theta_\mu|$ remains below a preset maximal step size $\delta\theta_\mathrm{max}$. We set $\delta\theta_\mathrm{max}=0.005$ for calculations with the Runge-Kutta method. We also note that in order to prevent potential numerical issues related to the inversion of the matrix $M$ in equation~(\ref{eq:ode}), we use the Tikhonov regularization approach, where a small number $\delta=10^{-6}$ is added to the diagonal of $M$. This stabilizes the matrix inversion when $M$ has a high condition number.

For benchmarking, we compare the AVQDS results with numerically exact data that are obtained using exact diagonalization via 
\begin{align}
|\Psi[t+\delta t]\rangle = e^{-i\delta t\hat{\mathcal{H}}[t]}|\Psi[t]\rangle\,,
\label{eq:ED}
\end{align}
on a uniformly discretized time mesh with a sufficiently small step size of $\delta t = 0.005$. It is important to note that all 1DCS and 2DCS results presented in this paper can be obtained with exact diagonalization, and the simulations are faster than AVQDS for the studied few-site high-spin models. However, exact diagonalization is generally limited to small systems due to exponential scaling with system size. The AVQDS approach is being developed as a near-term quantum algorithm, which leverages the scalable quantum computing power for unitary state evolution in the exponentially growing Hilbert space. The observables and matrix elements governing the equation of motion of the variational parameters, equation~\eqref{eq:ode}, are to be measured on quantum computers. A crucial first step is to investigate the required near-term quantum resources for specific applications, which are characterized by the number of CNOT gates in the quantum circuit. Simulations using statevector can provide important information to address this question, which is the focus of this paper.

\subsection{Initial state preparation}
The AVQDS simulation starts with the ground state, $\ket{\Psi_0}$, of the unperturbed Hamiltonian $\hat{\mathcal{H}}_0$ [see equation~\eqref{eq:Ham} with $\mathbf{B}=0$]. NISQ-friendly quantum algorithms for ground state preparation have been developed, including the variational quantum eigensolver (VQE)~\cite{vqe_theory, hardware_efficient_vqe}, quantum imaginary time evolution~\cite{qite_chan20}, and their variants~\cite{grimsleyAdaptiveVariationalAlgorithm2019,MayhallQubitAVQE,anastasiou2022,smqite,VQITE,AVQITE}. Here we adopt the qubit-ADAPT VQE method~\cite{MayhallQubitAVQE} in sections~\ref{sec:1Dvs2D}, \ref{sec:S-dep}, and \ref{sec:4-site} and the adaptive variational quantum imaginary time evolution (AVQITE) approach~\cite{AVQITE} in section~\ref{sec:N-dep}.
Starting with an easily prepared reference product state $\ket{\Phi_0}$, the qubit-ADAPT method constructs a problem-specific ansatz by appending a parametrized unitary each iteration with a subsequent reoptimization of all the parameters. The generator of the unitary that minimizes the energy gradient is chosen from a predefined operator pool. New unitaries are added to the ansatz until convergence in expectation value of $\hat{\mathcal{H}}_0$ is reached. Therefore, the ground state ansatz has the same pseudo-Trotter form~\eqref{eq:ansatz} and can be easily combined with AVQDS for the dynamics simulations. As a result, the initial AVQDS ansatz already has a finite number of variational parameters. 

In the AVQITE approach~\cite{AVQITE}, the McLachlan distance between the actual imaginary time-evolved state and the variational ansatz state is minimized. Similar to the AVQDS method, the McLachlan distance is kept below a specified threshold throughout the simulated time evolution by dynamically adding parameterized unitaries selected from a predefined operator pool to the variational ansatz. The resulting ground state ansatz also takes the pseudo-Trotter form~\eqref{eq:ansatz}.

In practice, we employ $\ket{\Phi_0} = \ket{0}^{\otimes N_\mathrm{q}}$ as our reference state for the ground state preparation using qubit-ADAPT VQE and AVQITE. Here $N_\mathrm{q}$ is the number of qubits for encoding the spin-$s$ Hamiltonian. We use the following operator pool for both ground state preparation with qubit-ADAPT VQE and real-time state propagation with AVQDS:
\begin{align}
    \mathcal{P}=&\{A_i: A \in \{\sigma^x,\sigma^y,\sigma^z\},\, 0 \le i < N_\mathrm{q} \}  \nonumber \\
    &\cup \{A_i B_j: A,B \in \{\sigma^x,\sigma^y,\sigma^z\},\, 0\le i < j < N_\mathrm{q}\}\,, 
    \label{eq:pool}
\end{align}
which includes all possible one- and two-qubit Pauli strings. To prepare the ground state of the quantum spin models in section~\ref{sec:N-dep},  we utilize the AVQITE method with the operator pool:
\begin{align}
\mathcal{P}=\{\sigma^y_i\}_{i=0}^{N_q-1}\cup \{\sigma^y_{i-1} \sigma^z_i\}_{i=1}^{N_q-1}\cup \{\sigma^z_{i-1} \sigma^y_i\}_{i=1}^{N_q-1}\,.
\label{eq:pool2}
\end{align}
For the corresponding dynamics simulation using the AVQDS approach, we utilize the Hamiltonian operator pool, which contains all individual terms of the Hamiltonian after transforming the high-spin operators of $\hat{\mathcal{H}}$ to qubits.

\begin{figure*}[t!]
\begin{center}
		\includegraphics[scale=0.7]{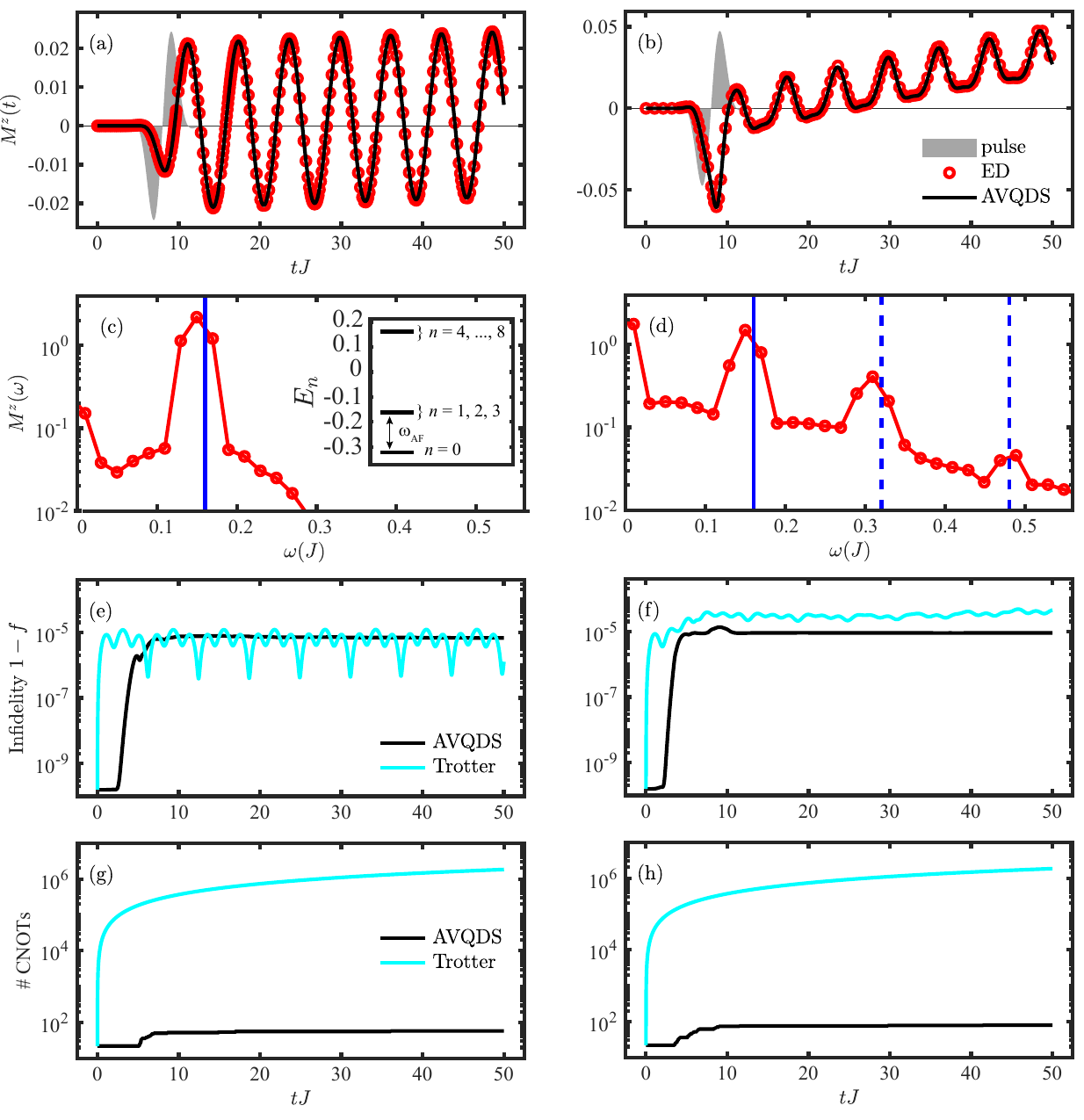}
		\caption{\textbf{One-dimensional coherent spectroscopy of a two-site spin-$s=1$ system.} Dynamics of the magnetization $M^z(t)$ induced by a single magnetic field pulse (shaded area) is shown for a $B$-field strength of (a) $B_0=0.5$ and (b) $B_0=3.0$. Model parameters are $D = 0.2$, $J = 1$, $K_a = K_b = 0$ such that $M^z(t=0)=0$. The result of the numerically exact simulations (red circles) are accurately reproduced by the results of the AVQDS simulations (black line). The corresponding spectra $M^z(\omega)$ in (c), (d) show second- and third harmonic generation peaks (vertical dashed lines) in addition to the fundamental harmonic peak at the antiferromagnetic magnon frequency $\omega=\omega_\mathrm{AF}$ (vertical solid line). 
  Inset of (c): Eigenenergies $E_n$ of the unperturbed Hamiltonian $\hat{\mathcal{H}}_0$.
  The Hamiltonian of the two-site spin-$s=1$ model has $(2s+1)^2=9$ eigenstates denoted by the index $n$. The energy difference between the ground state ($n=0$) and the first excited state ($n=1$) determines the magnon frequency $\omega_\mathrm{AF}=E_1-E_0\approx 0.16$, which is indicated by the double arrow. Note that the the states $n=1,2,3$ as well as $n=4,\cdots, 8$ are almost degenerate.
  (e), (f) Infidelities $1-f$ between AVQDS and exact state stay below $10^{-5}$ for most of the time evolution, which confirms the high accuracy of AVQDS. The first-order Trotter approach (cyan line) produces a comparable infidelity for the used uniform time step of $\delta t = 5 \times 10^{-3}$ in the simulations. 
  (g), (h) Required NISQ quantum resources measured by the number of CNOT gates. The number of CNOTs increases from initially 22 to 58 for low (g) and to 80 for strong $B$-field excitation (h). For the first-order Trotter method (cyan line), the number of CNOTs grows to $1.84\times 10^6$ at the final simulation of $t J =50$ for both driving magnetic fields of $B_0=0.5$ and $B_0=3.0$.}
		\label{fig2} 
\end{center}
\end{figure*}

\section{Comparison between 1DCS and 2DCS: physics and quantum resources \label{sec:1Dvs2D}}
To present the advantage of 2DCS in characterizing quantum high-spin models and to discuss the required quantum resources of the AVQDS approach, we consider a two-site spin-$s=1$ model. In this section, we neglect the single-ion anisotropy in equation~(\ref{eq:Ham}), i.~e., $K_\mathrm{a}=K_\mathrm{c}=0$, for simplicity. The energy unit is defined by setting the coupling constant $J$ to one while we consider a DM strength of $D=0.2$. The energy levels $E_n$ of the studied 2-site spin-$s=1$ model are presented in the inset of Fig.\figref{fig2}{(c)}.  The energy difference between the ground state ($n=0$) and the first excited state ($n=1$) determines the magnon frequency $\omega_\mathrm{AF}=E_1-E_0\approx 0.16$ which is indicated by the double arrow. The eigenenergies of the states $n=1,2,3$ as well as $E_n$ of the states $n=4,\cdots, 8$ are nearly degenerate. Note that the energy difference between $E_{n=1,2,3}$ and $E_{n=4,\cdots, 8}$ is approximately given by $2\omega_\mathrm{AF}$ such that the two-site spin-$s=1$ model exhibits a quasi-harmonic energy spectrum as illustrated below. To excite the quantum-spin system, we use magnetic field pulses of the form
\begin{align}
    B(t)=B_0\sin\left[\omega_0(t-t_0)\right]\,e^{-(t-t_0)^2/\Delta t^2}\,,
\end{align}                  
where $\omega_0$ denotes the center frequency of the pulse, $\Delta t$ determines the pulse duration, $t_0$ defines the pulse center in time domain while $B_0$ determines the magnetic field strength. In the simulations we set $\Delta t = 2$, $t_0 = 5$, and $\omega_0=1$ such that the magnon mode of the antiferromagnetic quantum spin system is resonantly driven by the magnetic field pulse.
 
\subsection{One-dimensional coherent spectroscopy}
Before studying 2DCS 
we first consider the quantum spin dynamics induced by a single magnetic field pulse $B(t)$. 
Figures\figref{fig2}{(a)} and\figref{fig2}{(b)} show the dynamics of the induced total magnetization along the $z$-direction (red circles), $M^z(t)$, after excitation with a single-cycle magnetic field pulse (shaded area) of strength $B_0=0.5$ and $B_0=3.0$, respectively. The quantum spin dynamics was obtained by solving equation~\eqref{eq:ED}. For low $B_0$, $M^z(t)$ shows harmonic oscillations after the magnetic field pulse excitation.  Compared to that, the quantum spin dynamics becomes anharmonic for strong magnetic field excitation (Fig.\figref{fig2}{(b)}). Next, we determine the frequencies contributing to $M^z(t)$ by applying a Fourier transformation. To obtain higher frequency resolution with minimal spectral leakage, we use the Blackman window function~\cite{Blackman}
\be   
W_\mathrm{B}(t) = 
     \begin{cases}
     0.42-0.5\,\cos(2\pi t/T_\mathrm{f}) \\ + \, 0.08\,\cos(4\pi t/T_\mathrm{f})\,, \quad \mathrm{for}\,\, t_1<t<t_2\\
     0\,,\qquad  \mathrm{otherwise}
    \end{cases}  \label{eq:bmwf}
\ee
within the Fourier transformation where $T_\mathrm{f}=t_2-t_1$. We set $t_1=0$ and $t_2 J=50$ for the calculations of the $M^z(\omega)$ spectra in this section. The resulting $M^z(\omega)$ spectra for $M^z(t)$ presented in Figs.\figref{fig2}{(a)} and\figref{fig2}{(b)} are plotted in Figs.\figref{fig2}{(c)} and\figref{fig2}{(d)}, respectively. For $B_0=0.5$ (Fig.\figref{fig2}{(c)}), $M^z(\omega)$ exhibits one pronounced peak at the antiferromagnetic magnon frequency $\omega_\mathrm{AF}\approx 0.16$ (vertical solid line). For high $B$-field driving of $B_0=3.0$ (Fig.\figref{fig2}{(d)}), $M^z(\omega)$  shows second harmonic (SHG) and third harmonic generation (3HG) peaks (vertical dashed lines) at frequencies  
$\omega_{\mathrm{SHG}}=2\omega_\mathrm{AF}$ and $\omega_{\mathrm{3HG}}=3\omega_\mathrm{AF}$, in addition to the fundamental harmonic signal at $\omega=\omega_\mathrm{AF}$. 

To identify the origin of the high harmonic generation peaks, we study the quantum spin dynamics after the magnetic field pulse where the time evolution of the system wavefunction is given by
\begin{align}
    |\Psi(t)\rangle=\sum_j a_j\,e^{-i E_j t}|\Psi_j\rangle\,.
\end{align}
Here, $a_j\equiv\langle\Psi_j|\Psi(t=0)\rangle$ while $|\Psi_j\rangle$ and $E_j$ are eigenfunctions and eigenenergies of the unperturbed Hamiltonian $\hat{\mathcal{H}}_0$, respectively. The dynamics of the magnetization can then be written as 
\begin{align}
    M^z(t)=\sum_{j,k} M^z_{j,k}\,a_j^* a_k e^{-i(E_k-E_j)t}\,.
\end{align}
Here $M^z_{j,k}\equiv\langle\Psi_j| \hat{S}^z|\Psi_k\rangle$ is the magnetic dipole matrix elements, which determines the strength of the transitions between the different eigenstates. The high-harmonic generation peaks in Figs.\figref{fig2}{(c)} and\figref{fig2}{(d)} result from resonant transitions between different eigenstates $|\Psi_j\rangle$ and $|\Psi_k\rangle$. Specifically, the energy position of the peaks in the $M^z(\omega)$ spectra corresponds to the difference between the energies of the eigenstates $|\Psi_j\rangle$ and $|\Psi_k\rangle$, i.~e., $\Delta E_{j,k}=\abs{E_j - E_k}$. An analysis of the magnetic dipole matrix elements $M^z_{j,k}$ reveals that the fundamental harmonic in Figs.\figref{fig2}{(c)} and\figref{fig2}{(d)} dominantly originates from a transition between the ground-state ($|\Psi_0\rangle$) and the first excited state ($|\Psi_1\rangle$). The second harmonic generation peak mainly results from a transition between the states $|\Psi_{3}\rangle$ and $|\Psi_{6}\rangle$, which follows from a multi-photon excitation process. Here, the magnetic field pulse first drives a transition between the ground state $|\Psi_0\rangle$ and $|\Psi_3\rangle$ which is followed by a transition from $|\Psi_{3}\rangle$ to $|\Psi_{6}\rangle$. The third harmonic generation peak harmonic generation peak dominantly stems from a transition between the states $|\Psi_{0}\rangle$ and $|\Psi_{6}\rangle$.
The peak close to $\omega\approx 0$ for high driving magnetic fields (Fig.\figref{fig2}{(d)}) stems from the excitation of eigenstates with comparable energy $E_{n}$. Specifically, the magnetic field resonantly drives transitions from the ground state to the excited states $|\Psi_{n=1,2,3}\rangle$. Here, the transition from the ground state to $|\Psi_{n=2}\rangle$ has a small dipole matrix element $M^z_{0,2}$ such that its contribution to the $M^z$ dynamics is negligible. The excitation of the states $|\Psi_{n=1,3}\rangle$ leads to a signal at frequency $E_{1,3}-E_0 \approx\omega_\mathrm{AF}$ and to a signal at the difference frequency $\Delta E_{3,1}=E_3-E_1\approx 0$ in the $M^z(\omega)$ spectra. The low frequency component $\Delta E_{3,1}$ is observable as a slow increase of $M^z(t)$ as a function of time in Fig.\figref{fig2}{(b)}.


\begin{figure}[t!]
\begin{center}
		\includegraphics[scale=0.5]{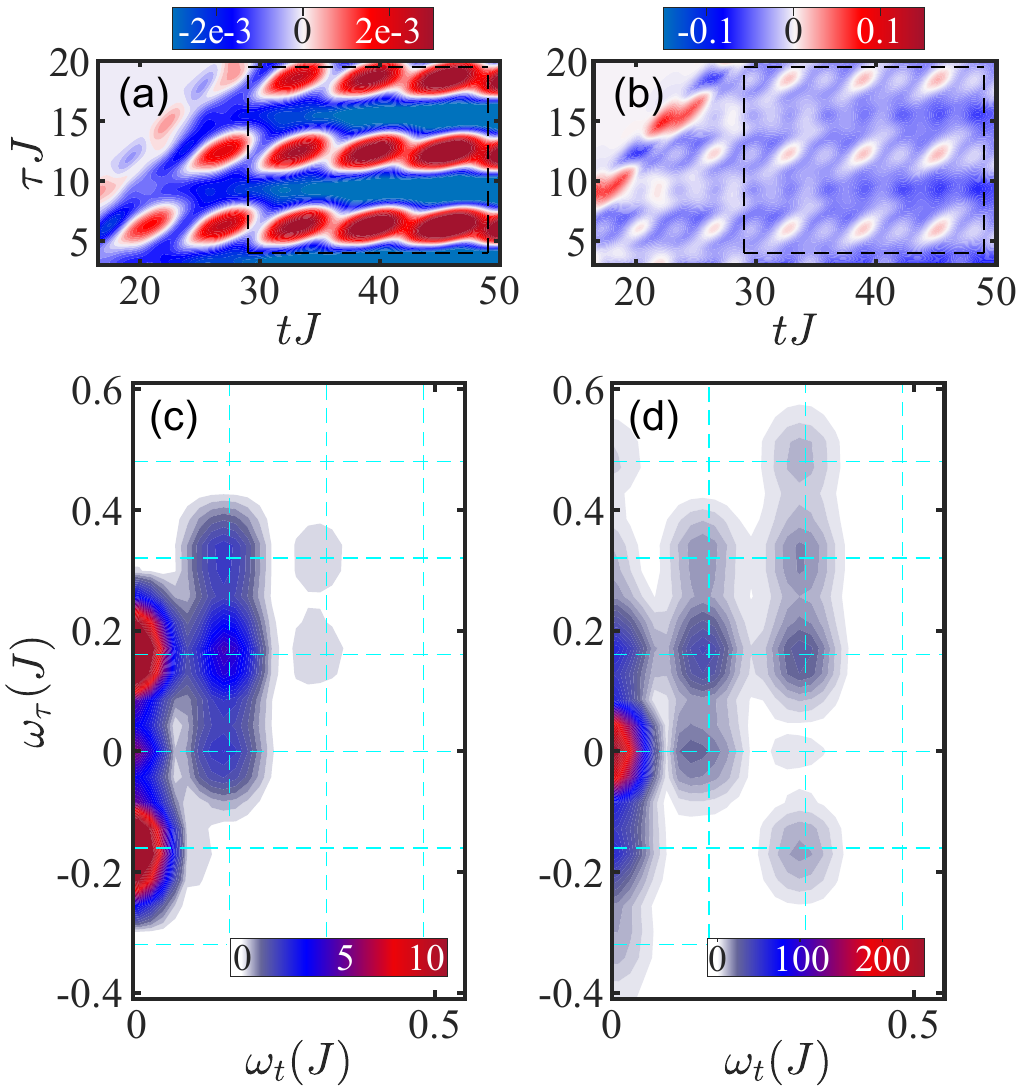}
		\caption{\textbf{Two-dimensional coherent spectroscopy of a two-site spin-$s=1$ system.} Nonlinear magnetization $M^z_\mathrm{NL}(t,\tau)$ as a function of time $t$ and inter-pulse delay $\tau$ is shown for a $B$-field strength of (a) $B_0=0.5$ and (b) $B_0=3.0$. Dashed rectangles indicate the region considered in the 2D Fourier transformation. (c), (d) Two-dimensional (2D) Fourier transform of $M^z_\mathrm{NL}(t,\tau)$ from (a), (b). Distinct peaks emerge at multiples of the magnon frequency $\omega_\mathrm{AF}$ in $\omega_t$- and $\omega_\tau$-directions indicated by dashed lines.}
		\label{fig3} 
\end{center}
\end{figure}

\subsection{AVQDS simulation results}
Next, we study the performance and required quantum resources of the AVQDS algorithm for the studied 2-site spin-$s=1$ model. To encode spin-$s=1$ operators into multi-qubit operators we use the Gray code in this section. A detailed comparison of AVQDS results with Gray code and standard binary encoding is given in section~\ref{sec:S-dep}. We use an operator pool of size $N_\mathrm{p}=66$ which contains all one- and two-qubit Pauli strings, equation~\eqref{eq:pool}. Here we note that even though the Hamiltonian~(\ref{eq:Ham}) after transformation to qubit operators contains Pauli terms up to size $p=4$ for $s=1$, the chosen operator pool only contains Pauli strings up to a weight of two. This is sufficient to accurately describe the quantum spin dynamics as illustrated below. Figures~\figref{fig2}{(a)} and\figref{fig2}{(b)} show the results of AVQDS (solid line) in addition to the exact simulation results (red circles). The AVQDS simulation accurately reproduces the exact results, with a maximal deviation $\max_t \abs{M^z_\mathrm{exact}(t)-M^z_\mathrm{AVQDS}(t)} = 2.5\times 10^{-4}$ ($1.2\times 10^{-3}$) for $B_0=0.5$ ($B_0=3.0$). To quantify the performance of the AVQDS approach, we plot in Figs.~\figref{fig2}{(e)} and \figref{fig2}{(f)} the corresponding infidelity, which is defined by $1-f=1-|\langle \Psi[\bth(t)]|\Psi_\mathrm{exact}(t)\rangle|^2$. Here, $|\Psi_\mathrm{exact}[t]\rangle$ represents the wavefunction obtained by solving equation~(\ref{eq:ED}), while $|\Psi[\bth(t)]\rangle$ is the variational wavefunction from AVQDS simulations. The infidelity stays below $10^{-5}$ for most of the time evolution even for strong magnetic field pulse excitation and at long simulation times, which demonstrates that the exact quantum spin dynamics is accurately reproduced by the AVQDS approach. 

The above AVQDS simulations are performed using statevector simulators, which assumes infinite sampling for measuring the matrix $M$~\eqref{eq:M} and vector $V$~\eqref{eq:V} without any hardware errors. Therefore, it is important to estimate the quantum resources needed for AVQDS calculations on NISQ devices. Here we focus on the quantum circuit complexity measured by the number of two-qubit CNOT gates. For simplicity, we assume all-to-all qubit connectivity (like on ion trap hardware) such that the implementation of the multi-qubit rotation gate $e^{-i\theta\hat{\A}}$ requires $2(p-1)$ CNOT gates for a Pauli string $\hat{\A}$ as the generator consisting of $p$ Pauli operators~\cite{nielsen2002quantum}. Figures\figref{fig2}{(g)} and \figref{fig2}{(h)} show the number of CNOTs as a function of time for the weak ($B_0=0.5$) and strong ($B_0=3.0$) magnetic field excitations studied in Figs.\figref{fig2}{(a)} and \figref{fig2}{(b)}. The initial-state preparation using the qubit-ADAPT-VQE method accurately reproduces the exact ground state with an infidelity of about $10^{-10}$. The resulting number of variational parameters $N_\theta$ of the initial state is 15. The corresponding variational ansatz~\eqref{eq:ansatz} has 4 single- and 11 two-qubit rotation gates, i.e., 22 CNOTs. During the time evolution the number of CNOTs mainly increases during the $B$-field driving, while only a minor increase is observed after the pulse excitation. In particular, the number of CNOTs increases from initially 22 to 58 for low and to 80 for strong $B$-field excitation at the final simulation time of $tJ=50$. The required quantum resource cost is also tied to the number of time steps $N_\mathrm{t}$, since one needs to measure $M$~\eqref{eq:M} and $V$~\eqref{eq:V} at each time step. For $B_0=0.5$ the number of time steps is $N_\mathrm{t}=3845$ and for $B_0=3.0$ it is $N_\mathrm{t}=7402$. 
The higher $N_\mathrm{t}$ for high $B$-field originates from the increased complexity of the dynamics due to the higher harmonic generations.

While we use the AVQDS approach to calculate the magnetization $M^z(t)$ in this paper, it can also be determined by other quantum computing approaches including Trotterization~\cite{lloyd1996, Trotter_dynamics_Knolle}. 
For comparison, we also calculated the $M^z(t)$ dynamics in Fig.~\ref{fig2} using the first-order Trotter approach. By expressing 
the Hamiltonian as $\hat{\mathcal{H}}=\sum_\mu\hat{h}_\mu[t]$ where $\hat{h}_\mu[t]$ is a 
Pauli string, the Trotterized dynamics follows from
\begin{align}
\label{eq:trotter}
	|\Psi[t+\delta t]\rangle = \Pi_\mu e^{i\delta t\,\hat{h}_\mu[t]}|\Psi[t]\rangle\,.
\end{align}  
Based on equation~(\ref{eq:trotter}), the circuit depth is determined by the number and weight (number of nonidentity Pauli operators in a Pauli string) of the Hamiltonian terms and grows linearly with the number of time steps. For the spin-$s=1$ model and Gray code studied in Fig.~\ref{fig2}, the Hamiltonian consists of 56 Pauli strings (Fig.~\figref{fig5}{a}). The resulting number of CNOT gates in the circuit increases by 184 during a single Trotter time step assuming full qubit-connectivity. To compare AVQDS and Trotter methods we use a uniform time step of $\delta t = 5 \times 10^{-3}$ in the Trotter simulation. This yields comparable accuracy as the AVQDS approach as illustrated in Figs.~\figref{fig2}{(e)} and \figref{fig2}{(f)}, where the infidelities of the first-order Trotter simulations (cyan lines) are shown alongside the corresponding infidelities of the AVQDS approach (black lines). However, the number of required CNOT gates for the first-order Trotter method, presented in Figs.~\figref{fig2}{(g)} and \figref{fig2}{(h)} as cyan lines, grows to $1.84\times 10^6$ at the final simulation time of $t J =50$ for both driving magnetic fields of $B_0=0.5$ and $B_0=3.0$. Consequently, the AVQDS approach requires four orders of magnitude fewer CNOT gates than the first-order Trotter dynamics simulations of comparable accuracy. This confirms that the AVQDS approach operates with much shallower quantum circuits than first-order Trotter method in agreement with previous studies~\cite{AVQDS}.

\begin{figure*}[t!]
\begin{center}
		\includegraphics[scale=0.45]{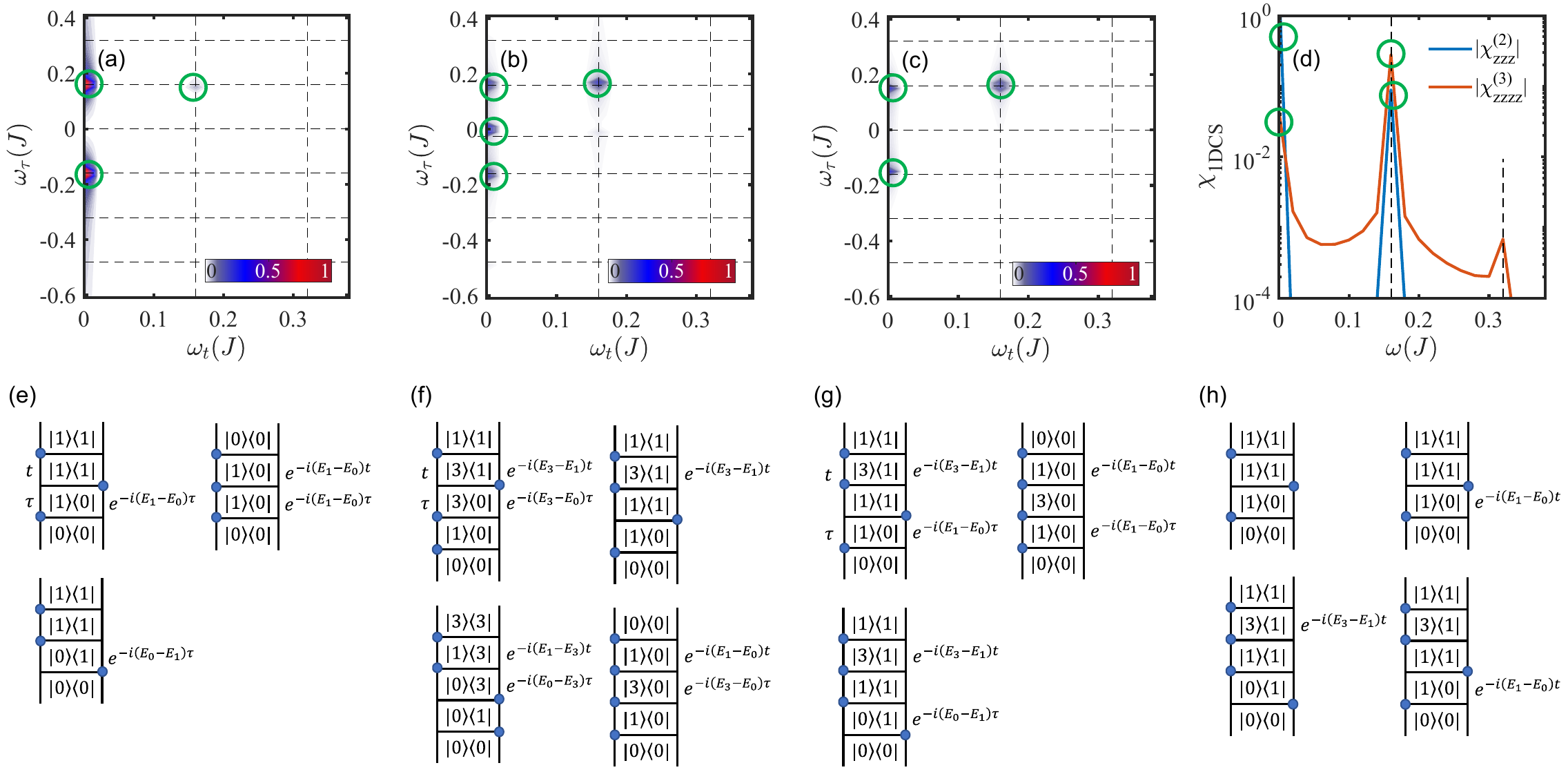}
		\caption{\textbf{Spectra of second-order and third-order susceptibilities: 1DCS versus 2DCS.} The two-dimensional spectrum of the second-order susceptibility $|\chi^{(2)}_{zzz}(\omega_t,\omega_\tau)|$ is shown in (a) while the corresponding spectra of third-order susceptibilities  $|\chi^{(3)}_{zzzz}(\omega_t,\omega_\tau,0)|$ and $|\chi^{(3)}_{zzzz}(\omega_t,0,\omega_\tau)|$ are presented in (b) and (c), respectively. (d) Spectra of the second-order, $|\chi^{(2)}_{zzz}(\omega)|$ (blue line),  and third-order, $|\chi^{(3)}_{zzzz}(\omega)|$ (red line), nonlinear susceptibilities of 1DCS. The dominant signals in (a)--(d) are indicated with circles. (e)–(h) Liouville pathways associated with the dominant signals in (a)–(d). The density matrix evolves from bottom to top. Eigenstates of $\hat{\mathcal{H}}_0$ are denoted by $|n\rangle$. 
        Operations of $\hat{S}^z$ on the bra and ket of the density matrix are denoted by circles. The exponentials indicate the phase accumulated during time intervals $t$ and $\tau$. We used explicit results of the matrix elements to identify the dominant terms. The location of the peaks is determined by the phase evolution during the time intervals $\tau$ and $t$, while the size of the peaks is given by the product of matrix elements.  size of the matrix elements. The density matrix evolves from bottom to top. }
		\label{fig4} 
\end{center}
\end{figure*}

\subsection{Two-dimensional coherent spectroscopy}

We now turn to 2DCS for characterizing quantum high-spin models. Therefore, we calculate the nonlinear magnetization along $z$-direction $M_{\text{NL}}^z(t)$~(\ref{eq:Mnl}) using the AVQDS approach for the two-pulse setup discussed in section~\ref{sec:model}. We compute $M^z_\mathrm{NL}(t,\tau)$ for $\tau J\in[3.5,20.0]$ using a delay-time stepping of $\Delta\tau J=0.1$. The time step in $t$ is determined adaptively to keep the maximal change of the variational parameters below $\delta \theta_{\text{max}} = 5.0\times 10^{-3}$, as discussed in Sec.~\ref{subsec:high_order_integrator}.
Figures\figref{fig3}{(a)} and \figref{fig3}{(b)} show $M^z_\mathrm{NL}(t,\tau)$, as a function of time $t$ and inter-pulse delay $\tau$ for low  and high $B$-field driving, respectively. The number of time steps required to obtain accurate simulation results depends on the delay time and ranges from 4174 to 12005 (from 6479 to 16120) for magnetic field strength of $B_0=0.5$ ($B_0=3.0$). Also the saturated number of CNOT gates changes with $\tau$. Up to 68 and 164 CNOT gates are needed to accurately simulate $M^z_\mathrm{NL}(t,\tau)$ in Figs.\figref{fig3}{(a)} and \figref{fig3}{(b)}, respectively.  

The nonlinear magnetization in Figs.\figref{fig3}{(a)} and \figref{fig3}{(b)} shows qualitative differences between low and high magnetic fields. At low $B$-field driving, the oscillations along $t$ and $\tau$ are dominated by the magnon frequency $\omega_\mathrm{AF}$, while new oscillation patterns become prominent at high $B_0$. To identify the different nonlinear contribution to $M^z_\mathrm{NL}(t,\tau)$, we perform a 2D Fourier transformation. Since the $t$-stepping is not equidistant due to the dynamical adjusted $\delta t$ in the AVQDS approach and also changes with $\tau$, we interpolate $M^z_\mathrm{NL}(t,\tau)$ to an equidistant $t$-stepping for all $\tau$ using linear interpolation before applying the 2D Fourier transformation. Since we are interested in probing the properties of the high spin-$s$ model, we apply the 2D Fourier transformation on the time region after both pulses have excited the quantum spin system, which is indicated by the dashed rectangles in Figs.\figref{fig3}{(a)} and \figref{fig3}{(b)}. This is done by applying the Blackman window function~\eqref{eq:bmwf} in both $t$- and $\tau$-direction, i.~e.~$W_\mathrm{B}(t,\tau)\equiv W_\mathrm{B}(t)W_\mathrm{B}(\tau)$, with $t_1 J=29.0$, $t_2 J=49.0$, $\tau_1 J=4.0$, and $\tau_2 J=19.5$. The resulting 2D spectra, $M^z_\mathrm{NL}(\omega_t,\omega_\tau)$,
are plotted in Figs.\figref{fig3}{(c)} and \figref{fig3}{(d)}. Multiple peaks are visible in the 2D spectra.
At low field $B_0=0.5$ in Fig.\figref{fig3}{(c)}, the 2DCS spectrum shows three dominant peaks at $(\omega_t,\omega_\tau)=(0,\pm\omega_\mathrm{AF})$, $(\omega_t,\omega_\tau)=(0,0)$, and $(\omega_t,\omega_\tau)=(\omega_\mathrm{AF},\omega_\mathrm{AF})$. Weaker signals are observable at $(\omega_\mathrm{AF},\omega_\mathrm{AF}\pm\omega_\mathrm{AF})$. Additional peaks are visible at $\omega_t=2\omega_\mathrm{AF}$. 
The signals at $\omega_t=2\omega_\mathrm{AF}$ become as strong as that at $\omega_t=\omega_\mathrm{AF}$ with high $B$-field driving (Fig.~\figref{fig3}{(d)}), which indicates that high-order nonlinearities dominate over lower ones at elevated $B_0$.

The different nonlinear processes contributing to the 2DCS spectra in Figs.\figref{fig3}{(c)} and \figref{fig3}{(d)} can be identified by applying a susceptibility expansion~\cite{mukamel1995principles,Wan2019,Nandkishore2021} of equation~\eqref{eq:Mnl}. Approximating the pulse shape by $\delta$-functions, the magnetic field becomes
\begin{align}
	B^z=A_0^z\,\delta(t) + A^z_\tau\,\delta(t-\tau)\,,
\end{align}	
where the first pulse is centered at time $t=0$ and the second one at $t=\tau$. $A$ denotes the pulse area. The laser-light induced nonlinear magnetization along $z$-direction measured at time $t+\tau$ can then be written as~\cite{Wan2019,Choi2020,Nandkishore2021}
\begin{align}
	&M_\mathrm{NL}^z(t+\tau)/N \nonumber \\ 
        &\equiv (M^z_\mathrm{12}(t+\tau)-M^z_\mathrm{1}(t+\tau)-M^z_\mathrm{2}(t+\tau))/N \nonumber \\
	&= \chi^{(2)}_{zzz}(t,\tau)\,A^z_\tau A^z_0 \nonumber \\
	&+\chi^{(3)}_{zzzz}(t,\tau,0)\,A^z_\tau (A^z_0)^2+\chi^{(3)}_{zzzz}(t,0,\tau)\,(A^z_\tau)^2 A^z_0\nonumber \\
	&+\mathcal{O}(B^4)\,,
 \label{eq:Mexp}
\end{align}
where $N$ denotes the number of spin sites. 
The second-order susceptibility is defined by~\cite{Wan2019,Choi2020}
\begin{align}
	\chi^{(2)}_{zzz}(t,\tau)=-\frac{1}{N}\theta(t)\theta(\tau)\langle\left[\left[M^z(t+\tau),M^z(\tau)\right],M^z(0)\right]\rangle\,.
\end{align}
After applying the Fourier transformation, 
\begin{align}
	&\chi^{(2)}_{\alpha\beta\beta}(\omega_t,\omega_\tau)=\int_{0}^\infty \mathrm{d}t\int_{0}^\infty \mathrm{d}\tau\,\chi^{(2)}_{\alpha\beta\beta}(t,\tau)e^{i\,\omega_t t}e^{i\,\omega_\tau\tau}\,,
\end{align}
and expanding the result in term of the eigenstates $|\Psi_n\rangle$ of the unperturbed Hamiltonian $\hat{\mathcal{H}}_0$, we obtain
\begin{align}
	&\chi^{(2)}_{zzz}(\omega_t,\omega_\tau)\nonumber \\ 
 &=\frac{1}{N}\sum_{\mu,\nu}S_{zzz}^{\mu,\nu}\left[L_{0,\mu}(\omega_t)L_{0,\nu}(\omega_\tau)-L_{\mu,\nu}(\omega_t)L_{0,\nu}(\omega_\tau)\right. \nonumber \\ &\qquad\qquad\qquad\left.-L_{\mu,\nu}(\omega_t)L_{\mu,0}(\omega_\tau)+L_{\nu,0}(\omega_t)L_{\mu,0}(\omega_\tau)\right]\,.
 \label{eq:chi2}
\end{align}
Here,
\begin{align}
	L_{\mu,\nu}(\omega)=\frac{1}{\omega+i 0^+ + E_\mu-E_\nu}
\end{align}
with infinitesimal positive quantity $0^+$ and eigenenergies $E_\mu$ of $\hat{\mathcal{H}}_0$ while 
\begin{align}
	S_{zzz}^{\mu,\nu}=M^z_{0,\mu}\,M^z_{\mu,\nu}\,M^z_{\nu,0}\,.
\end{align}
The second-order susceptibility vanishes for inversion-symmetric spin systems. In the model~\eqref{eq:Ham}, it becomes finite due to the presence of inversion-symmetry breaking DM interaction.

The third-order susceptibilities are defined by~\cite{Nandkishore2021}
\begin{align}
	&\chi^{(3)}_{zzzz}(t,\tau,0) \nonumber \\
 &=-\frac{i}{N}\theta(t)\theta(\tau)\langle\left[\left[\left[M^z(t+\tau),M^z(\tau)\right],M^z(0)\right],M^z(0)\right]\rangle\,, \nonumber \\
	&\chi^{(3)}_{zzzz}(t,0,\tau) \nonumber \\
 &=-\frac{i}{N}\theta(t)\theta(\tau)\langle\left[\left[\left[M^z(t+\tau),M^z(\tau)\right],M^z(\tau)\right],M^z(0)\right]\rangle\,.
\end{align}
After applying the 2D Fourier transformation, we find
\begin{align}
	&\chi^{(3)}_{zzzz}(\omega_t,\omega_\tau,0)\nonumber \\
 &=\frac{i}{N}\sum_{\mu,\nu,\lambda}S^{\mu,\nu,\lambda}_{zzzz}\left[L_{0,\mu}(\omega_t)L_{0,\nu}(\omega_\tau)-L_{\mu,\nu}(\omega_t)L_{0,\nu}(\omega_\tau)\right. \nonumber \\ &\qquad\qquad\qquad\left.
 -2 L_{\mu,\nu}(\omega_t)L_{\mu,\lambda}(\omega_\tau)+2 L_{\nu,\lambda}(\omega_t)L_{\mu,\lambda}(\omega_\tau)\right. \nonumber \\ &\qquad\qquad\qquad\left.+L_{\nu,\lambda}(\omega_t)L_{\nu,0}(\omega_\tau)-L_{\lambda,0}(\omega_t)L_{\nu,0}(\omega_\tau)\right]\,, \nonumber \\
	&\chi^{(3)}_{zzzz}(\omega_t,0,\omega_\tau)\nonumber \\
 &=\frac{i}{N}\sum_{\mu,\nu,\lambda}S^{\mu,\nu,\lambda}_{zzzz}\left[L_{0,\mu}(\omega_t)L_{0,\lambda}(\omega_\tau)-2 L_{\mu,\nu}(\omega_t)L_{0,\lambda}(\omega_\tau)\right. \nonumber \\ &\qquad\qquad\qquad\left.+ L_{\nu,\lambda}(\omega_t)L_{0,\lambda}(\omega_\tau)- L_{\mu,\nu}(\omega_t)L_{\mu,0}(\omega_\tau)\right. \nonumber \\ &\qquad\qquad\qquad\left.+2 L_{\nu,\lambda}(\omega_t)L_{\mu,0}(\omega_\tau)-L_{\lambda,0}(\omega_t)L_{\mu,0}(\omega_\tau)\right]\,,
 \label{eq:chi3}
\end{align}
where 
\begin{align}	
S^{\mu,\nu,\lambda}_{zzzz}\equiv M^z_{0,\mu}M^z_{\mu,\nu}M^z_{\nu,\lambda}M^z_{\lambda,0}\,.
\end{align}
Based on equations~\eqref{eq:chi2} and \eqref{eq:chi3} the peaks in the 2DCS spectra in Figs.\figref{fig3}{(c)} and \figref{fig3}{(d)} emerge at energies corresponding to energy differences $E_\mu-E_\nu$ in both $\omega_t$ and $\omega_\tau$-directions. As a result, the energy positions along $\omega_t$ and $\omega_\tau$ characterize two transitions between different eigenstates. The spectral weight of the peaks in the 2DCS spectra is determined by $S_{zzz}^{\mu,\nu}$ for second-order and $S^{\mu,\nu,\lambda}_{zzzz}$ for third-order susceptibilities and the number of transitions contributing to the signals.

Figures\figref{fig4}{(a-c)} show the 2D spectra of the second-order, $|\chi^{(2)}_{zzz}(\omega_t,\omega_\tau)|$, and  third-order susceptibilities,  $|\chi^{(3)}_{zzzz}(\omega_t,\omega_\tau,0)|$ and $|\chi^{(3)}_{zzzz}(\omega_t,0,\omega_\tau)|$, respectively. The second-order susceptibility is about one order of magnitude larger than the third-order susceptibilities for the model parameters used in this work. The second-order susceptibility shows two dominant peaks at $(\omega_t,\omega_\tau)=(0,\pm\omega_\mathrm{AF})$ and a weaker peak at $(\omega_t,\omega_\tau)=(\omega_\mathrm{AF},\omega_\mathrm{AF})$ in agreement with the 2DCS spectrum presented in Fig.\figref{fig3}{(c)}. The Liouville pathways for the nonlinear processes leading to the these peaks are presented Fig.\figref{fig4}{(e)}. The two dominant peaks in Fig.\figref{fig4}{(a)} originate from a transition between eigenstates $|\Psi_0\rangle$ and $|\Psi_1\rangle$ in $\omega_\tau$-direction while the signal at $(\omega_t,\omega_\tau)=(\omega_\mathrm{AF},\omega_\mathrm{AF})$ results from a transition between eigenstates $|\Psi_0\rangle$ and $|\Psi_1\rangle$ in $\omega_t$- as well as in $\omega_\tau$-direction. Compared to that, the third-order susceptibilities in Figs.\figref{fig4}{(b)} and \figref{fig4}{(c)} show four dominant peaks at $(\omega_t,\omega_\tau)=(0,\pm\omega_\mathrm{AF})$, $(\omega_t,\omega_\tau)=(0,0)$, and $(\omega_t,\omega_\tau)=(\omega_\mathrm{AF},\omega_\mathrm{AF})$. The Liouville pathways associated with the peaks are illustrated in Figs.\figref{fig4}{(f)} and \figref{fig4}{(g)}. The additional peaks present in Fig.\figref{fig3}{(c)} need to be described by susceptibilities beyond third order. In particular, the 2DCS spectrum at high $B_0$ in Fig.\figref{fig3}{(d)} demonstrates that higher order susceptibilities dominate over lower ones at elevated $B$-field driving, since the strong signals at $\omega_t=2\omega_\mathrm{AF}$ are not covered by the third-order susceptibilities in Figs.\figref{fig4}{(b)} and \figref{fig4}{(c)}. 

To directly compare 1DCS and 2DCS, we also calculate the susceptibilities of 1DCS. Here, the system is excited by a single pulse polarized along the $z$-direction
\begin{align}
	B^z=A_0^z\,\delta(t)\,.
\end{align}	
The laser-driven magnetization along $z$-direction then reads~\cite{Wan2019,Nandkishore2021,Choi2020}
\begin{align}
	M^z(t)/N&=\chi^{(1)}_{zz}(t)A_0^z+\chi^{(2)}_{zzz}(t)(A_0^z)^2+\chi^{(3)}_{zzzz}(t)(A_0^z)^3 \nonumber \\
 &+\mathcal{O}(B^4)\,,
\end{align} 
with linear, second-order, and third-order susceptibilities defined by
\begin{align}
	&\chi^{(1)}_{zz}(t)=\frac{i}{N}\theta(t)\langle\left[M^z(t),M^z(0)\right]\rangle\,,\nonumber \\
    &\chi^{(2)}_{zzz}(t)=-\frac{1}{N}\theta(t)\langle\left[\left[M^z(t),M^z(0)\right],M^z(0)\right]\rangle\,, \nonumber \\
    &\chi^{(3)}_{zzzz}(t)=-\frac{i}{N}\theta(t)\langle\left[\left[\left[M^z(t),M^z(0)\right],M^z(0)\right],M^z(0)\right]\rangle\,.
\end{align}
After applying the Fourier transformation, the susceptibilities become 
\begin{align}
	&\chi_{zz}^{(1)}(\omega)=\frac{1}{N}\sum_\mu S_{zz}^\mu\left[L_{0,\mu}(\omega)-L_{\mu,0}(\omega)\right]\,,\nonumber \\
 &\chi^{(2)}_{zzz}(\omega)=\frac{i}{N}\sum_{\mu,\nu}S_{zzz}^{\mu,\nu}\left[L_{0,\mu}(\omega)-2 L_{\mu,\nu}(\omega)+L_{\nu,0}(\omega)\right]\,, \nonumber \\
 &\chi^{(3)}_{zzzz}(\omega)=-\frac{\mathrm{1}}{N}\sum_{\mu,\nu,\lambda}S^{\mu,\nu,\lambda}_{zzzz}\left[L_{0,\mu}(\omega)-3L_{\mu,\nu}(\omega)\right. \nonumber \\
 &\qquad\qquad\qquad\left. +3 L_{\nu,\lambda}(\omega)-L_{\lambda,0}(\omega)\right]\,,
\end{align}
where
\begin{align}
	S_{zz}^{\mu}\equiv M^z_{0,\mu}\,M^z_{\mu,0}\,.
\end{align}
Figure\figref{fig4}{(d)} shows the spectra of the second- and third-order nonlinear susceptibilities of 1DCS. The second-order susceptibility exceeds $\chi^{(3)}_{zzzz}(\omega)$ in agreement with Figs.\figref{fig4}{(a-c)}. Both susceptibilities exhibit two pronounced peaks at $\omega=0$ and $\omega=\omega_\mathrm{AF}$. The 
Liouville pathways for these peaks in Fig.\figref{fig4}{(h)} show that the signals stem from transitions between eigenstates $|\Psi_0\rangle$ and $|\Psi_1\rangle$ as well as $|\Psi_1\rangle$ and $|\Psi_3\rangle$. As a result, 2DCS resolves more transitions between eigenstates. In particular, while the transitions along $\omega_t$ agree with the transitions observed with 1DCS, new transitions show up along $\omega_\tau$-directions. For example, the transition between eigenstates $|\Psi_0\rangle$ and $|\Psi_3\rangle$ (Fig.\figref{fig4}{(f)}) only shows up in $\omega_\tau$-direction and thus cannot be detected with 1DCS. Specifically, the transition between eigenstates $|\Psi_0\rangle$ and $|\Psi_3\rangle$ contributes to the same peak as the transition from $|\Psi_0\rangle$ to $|\Psi_1\rangle$ in 1DCS. 
However, since the matrix element of $M^z_{1,0}$ is about one order of magnitude larger than $M^z_{3,0}$, the transition from 
$|\Psi_0\rangle$ to $|\Psi_3\rangle$ is masked in 1DCS. In 2DCS it becomes
observable via the two-photon excitation process from $|\Psi_0\rangle$ to $|\Psi_1\rangle$ followed by an excitation from $|\Psi_1\rangle$ to $|\Psi_3\rangle$, which is combined with the transition from $|\Psi_0\rangle$ to $|\Psi_1\rangle$ in $\omega_t$-direction based on the Liouville pathway in Fig.\figref{fig4}{(f)}. As a general trend,  when studying high-order susceptibilities and higher spin-$s$ models, 2DCS is capable to resolve more transitions between eigenstates than 1DCS, which demonstrates that 2DCS provides higher resolution of the energy states of high-spin models.

\begin{figure}[t!]
\begin{center}
		\includegraphics[scale=0.60]{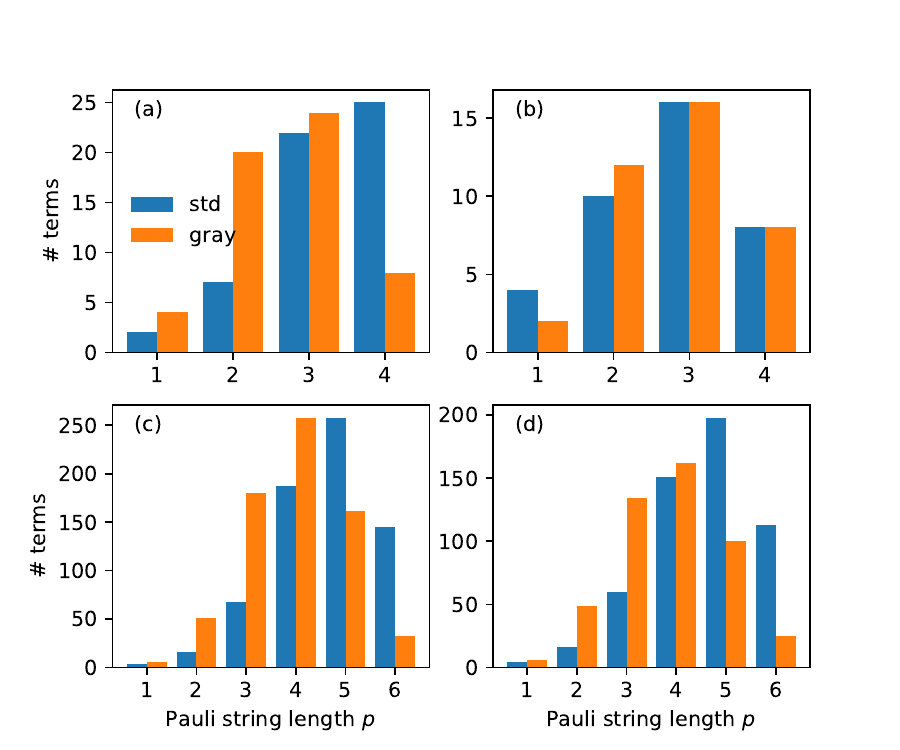}
		\caption{\textbf{Complexity of qubit Hamiltonian for different spin magnitudes $s$.}
  Number of Hamiltonian terms with Pauli string length $p$ 
for (a) $s=1$, (b) $s=3/2$, (c) $s=2$, and (d) $s=5/2$. The standard binary encoding (std, blue bars) is compared with the Gray code (orange bars).}
		\label{fig5} 
\end{center}
\end{figure}

\begin{figure*}[t!]
\begin{center}
		\includegraphics[scale=0.75]{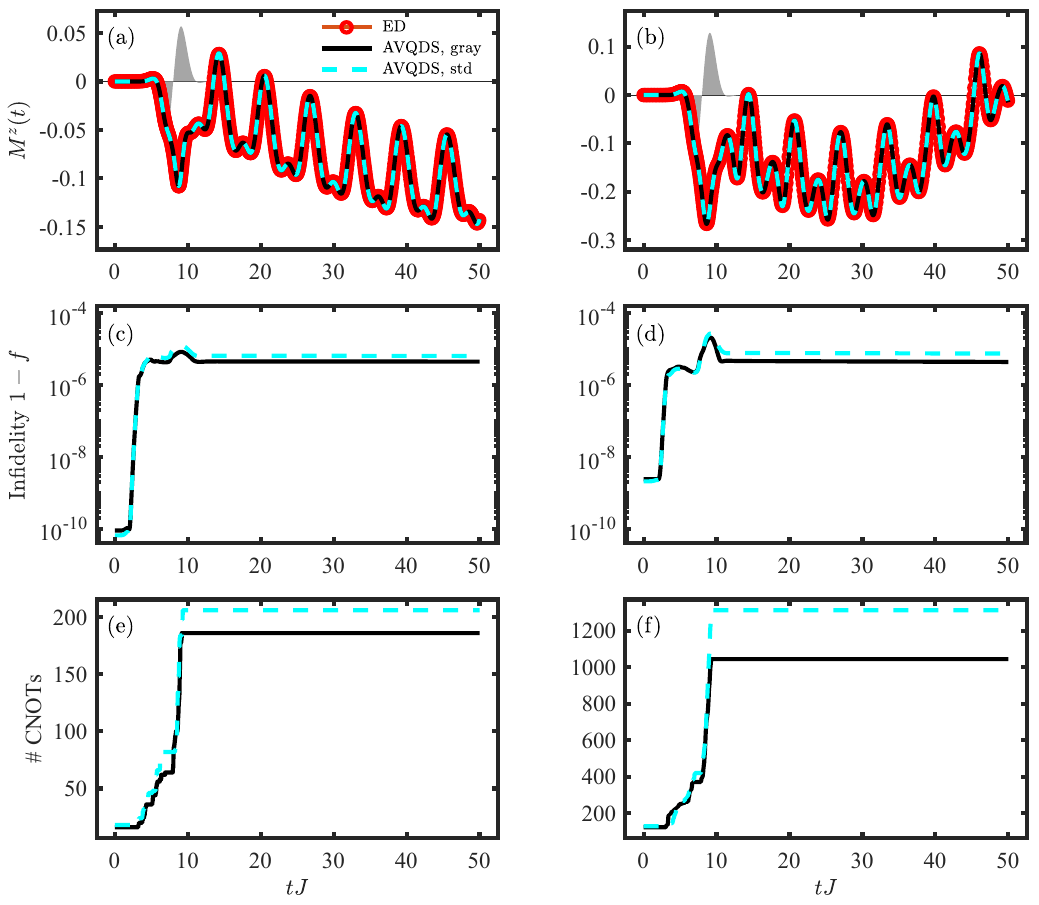}
		\caption{\textbf{Magnetic field induced quantum spin dynamics of two-site spin-$s=3/2$ and $s=5/2$ models for 1DCS.} Dynamics of the magnetization $M^z(t)$ induced by a single magnetic field pulse (shaded area) is presented for the quantum spin model with (a) $s=3/2$ and (b) $s=5/2$ using a magnetic field strength of $B_0=4.5$. The result of the exact simulations (red circles) is compared with the results of the AVQDS approach where the Gray code (solid black line) and standard binary encoding (std, dashed cyan line) was used. The corresponding infidelities in (c) and (d) demonstrate the high accuracy of the AVQDS simulations with a fidelity of at least $99.99\%$. The number of CNOT gates in (e) and (f) only increases during the $B$-field driving and is already saturated after a simulation time of $tJ=11$. The Gray code requires less CNOT gates with a saturated number of 186 for $s=3/2$ and 1044 for $s=5/2$ compared to 206 for $s=3/2$ and 1312 for $s=5/2$ for the standard binary encoding.}
		\label{fig6} 
\end{center}
\end{figure*}

\section{Quantum resource scaling with spin magnitude \label{sec:S-dep}}

For the purpose of preparing 2DCS calculations of high-spin models on quantum devices, it is important to deduce how the quantum resources vary with spin $s$, in a range of about $[1/2, 5/2]$, as the quantum fluctuations are proportional to $1/s$~\cite{auerbach1998interacting}. We estimate the required quantum resources for AVQDS using two different high-$s$ encodings introduced in section~\ref{sec:S_trans}, i.~e.,~standard binary encoding and Gray code. Here, we focus on the required quantum resources for 1DCS for simplicity. Note that 2DCS requires up to twice as many quantum resources compared to 1DCS, as discussed in section~\ref{sec:1Dvs2D}. Figure~\ref{fig5} shows histograms of the number of Pauli string of length $p$ in representing the Hamiltonian
for (a) $s=1$, (b) $s=3/2$, (c) $s=2$, and (d) $s=5/2$. The standard binary encoding (blue bars) is compared with the Gray code (orange bars). The transformation of $s=1$ and $s=3/2$ spin operators to qubit operators requires 2 qubits, while 3 qubits are needed to encode $s=2$ and $s=5/2$ spin operators. Thus, encoded $s=1$ and $s=3/2$ Hamiltonians contain Pauli strings up to weight $p=4$ and the $s=2$ and $s=5/2$ Hamiltonians up to weight $p=6$. The standard binary and Gray encodings have roughly the same total number of Hamiltonian terms. However, the number of longer Pauli strings in the Hamiltonian is reduced for the Gray encoding, which suggests that the Gray code may require less quantum resources compared to the standard binary encoding.

Figures~\figref{fig6}{(a)} and \figref{fig6}{(b)} present the magnetization dynamics $M^z(t)$ obtained for spin-$s=3/2$ and $s=5/2$ models, respectively, using a single strong magnetic field pulse (shaded area) with amplitude $B_0=4.5$. The exact simulation results (red circles) obtained by solving equation~\eqref{eq:ED} are compared with AVQDS results using the Gray encoding (solid black line) and the standard binary encoding (dashed cyan line). The AVQDS results closely match the exact simulation results with high accuracy for both studied high-spin models. To provide a more detailed measure of the accuracy of AVQDS, the corresponding infidelities are plotted in Figs.\figref{fig6}{(c)} and \figref{fig6}{(d)}. The infidelity stays below $10^{-4}$ throughout the time evolution for both models. In particular, the infidelity remains constant after the pulse excitation for both encodings which demonstrates that AVQDS produces accurate quantum spin dynamics even over long simulation times. 

\begin{figure*}[t!]
\begin{center}
		\includegraphics[scale=0.80]{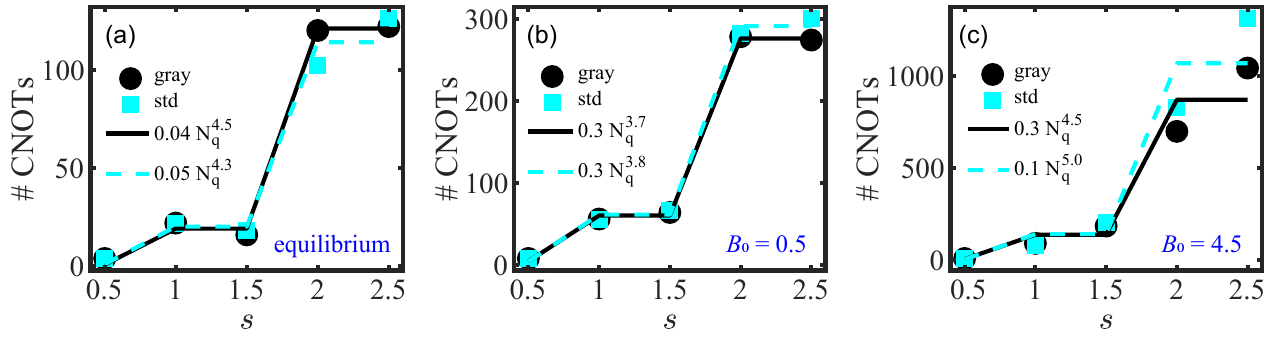}
		\caption{\textbf{Initial and saturated number of CNOT gates as a function of spin $s$.} (a) Required number of CNOT gates for the initial-state preparation as a function of spin $s$ for Gray code (black circles) and standard binary encoding (cyan squares). Fittings using a power-law function of the form $\alpha N_\mathrm{q}^\beta$ are shown as solid black and dashed cyan lines for Gray code and standard binary encoding, respectively. For $s=1/2$, $N_\mathrm{q}=2$ while for $s=1$ and $s=3/2$ ($s=2$ and $s=5/2$) the number of qubits is $N_\mathrm{q}=4$ ($N_\mathrm{q}=6$). Both encodings require comparable number of CNOT gates to prepare the initial state. The corresponding results for the saturated number of CNOT gates are shown in (b) and (c) for weak ($B_0=0.5$) and strong ($B_0=4.5$) $B$-field excitation. For low $B$-field excitation, the saturated number of CNOT gates is similar for both encodings, while at high $B$-field driving the Gray code requires less CNOT gates for high spins $s$.}
		\label{fig7} 
\end{center}
\end{figure*}

The required quantum resources in terms of CNOT gates are shown in Figs.~\figref{fig6}{(e)} and \figref{fig6}{(f)}. We use the qubit-ADAPT-VQE method to prepare the ground state with a final infidelity of about $10^{-10}$ ($10^{-9}$) for $s=3/2$ ($s=5/2$) models. Both encodings require a comparable number of CNOT gates for the initial state preparation.
During the time evolution, the number of CNOT gates only increases during the $B$-field driving and saturates already after a simulation time of $tJ=11$. The saturated number of CNOTs is lower for the Gray encoding, consistent with the simpler Hamiltonian representation in Gray encoding than that in standard binary encoding observed in Fig.~\ref{fig5}. 

Figure~\figref{fig7}{(a)} summarizes the $s$-dependence of the required quantum resources for the initial-state preparation. The result of Gray code (black circles) is compared with the result of the standard binary encoding  (cyan squares). Both encodings require a comparable number of CNOT gates. A fitting using a power-law function of the form $\alpha N_\mathrm{q}^\beta$, where $N_\mathrm{q}$ denotes the number of qubits needed to encode the quantum-spin Hamiltonian~\eqref{eq:Ham}, yields $\beta\approx 4.5$ (solid black line) and $\beta\approx 4.3$ (dashed cyan line) for the Gray code and standard binary encoding, respectively. 
The corresponding $s$-dependence of the saturated number of CNOT gates is presented in Figs.~\figref{fig7}{(b)} and \figref{fig7}{(c)} for a small ($B_0=0.5$) and large ($B_0=4.5$) $B$-field strength. For low $B$-field amplitude, the saturated number of CNOT gates is similar for both encodings. A power-law function fit produces $\beta\approx 3.7$ ($\beta\approx 3.8$) for Gray code (standard binary encoding). However, at high $B$-field driving, the Gray code requires less CNOT gates. The fitting leads to a scaling of approximately $0.3 N_\mathrm{q}^{4.5}$ and $0.1 N_\mathrm{q}^{5.0}$ for the Gray code and standard binary encoding, respectively. 
Since the Gray code requires less quantum resources than standard binary encoding, we use this encoding for the application in next section.

\section{Quantum resource scaling with system size \label{sec:N-dep}}
In this section, we study how the required quantum resources scale with the number of sites $N$ for spin $s=1/2$, $s=1$, and $s=3/2$ systems. To encode the spin-$s$ operators into multi-qubit operators, we employ the Gray encoding, which requires slightly fewer quantum resources compared to the standard binary encoding as demonstrated in section~\ref{sec:S-dep}. To prepare the ground state of the quantum spin models, we utilize the AVQITE method with the operator pool~\eqref{eq:pool2}, while we use the Hamiltonian operator pool, which is composed of all the individual Pauli strings in the Hamiltonian, for the quantum dynamics simulations using the AVQDS approach. 

\begin{figure*}[t!]
\begin{center}
		\includegraphics[scale=0.59]{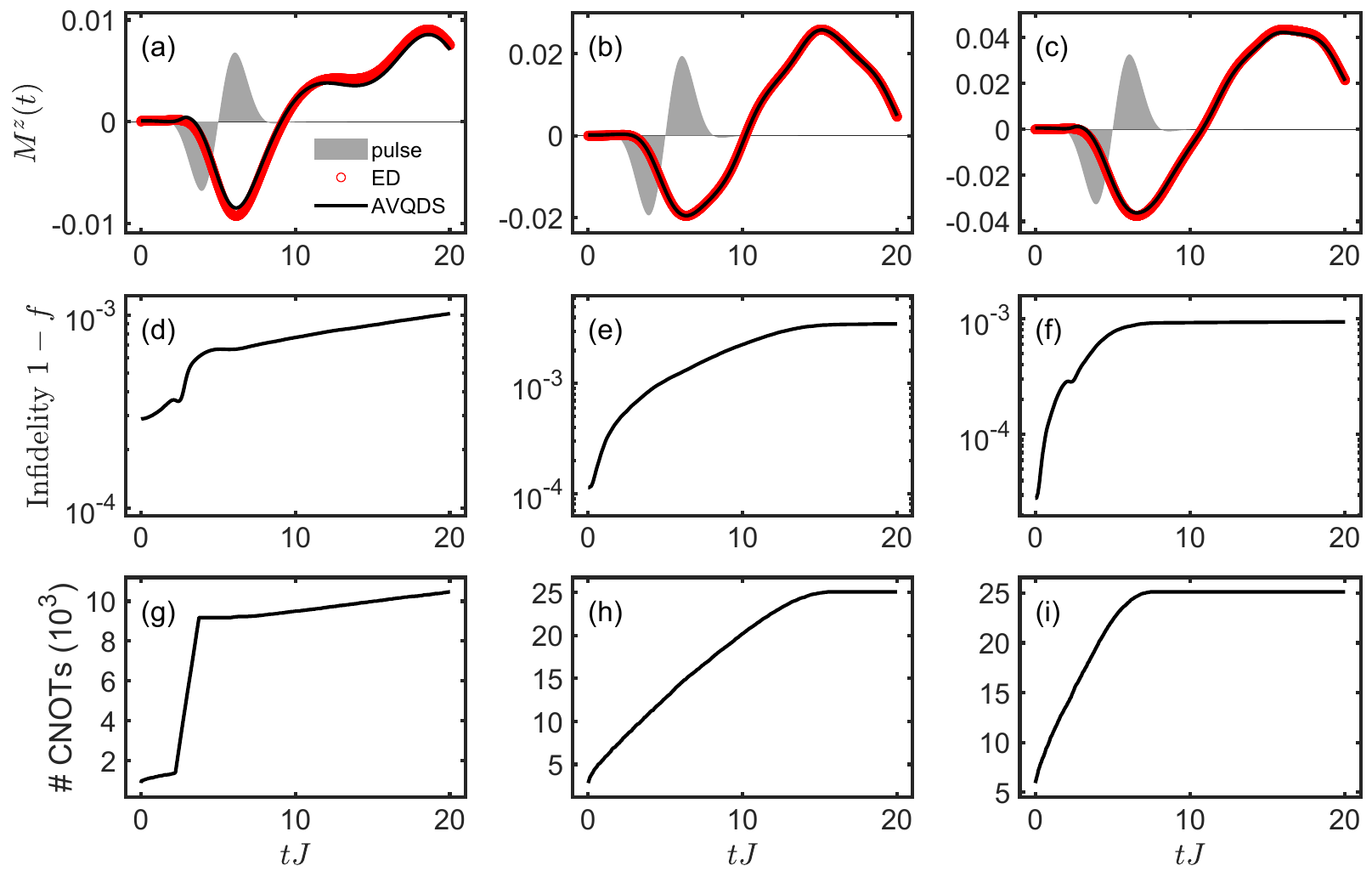}
		\caption{\textbf{Magnetic field induced quantum spin dynamics of 12-site spin-$s=1/2$, 6-site spin-$s=1$, and 6-site spin-$s=3/2$ models for 1DCS.} 
		The dynamics of the magnetization $M^z(t)$ induced by a single magnetic field pulse of strength $B_0=0.5$ (shaded area) are presented for (a) 12-site spin-$s=1/2$, (b) 6-site spin-$s=1$, and (c) 6-site spin-$s=3/2$ models. The results of the exact simulations are denoted by red circles, while those of the AVQDS approach (utilizing the Gray code) are shown as solid black lines. The corresponding infidelities in (d)--(f) demonstrate the high accuracy of the AVQDS simulations, with an infidelity smaller than $1.1\times 10^{-3}$ throughout the time evolution. The required number of CNOT gates, shown in (g)--(i), primarily increases during the $B$-field driving before saturating after a simulation time of about $tJ=15$ for $s=1$ and $tJ=7$ for $s=3/2$, while it shows a slow increase after the pulse for $s=1/2$. The saturated number of CNOTs is 10468, 25066, and 25118 for the 12-site spin-$s=1/2$, 6-site spin-$s=1$, and 6-site spin-$s=3/2$ models, respectively.}
		\label{fig9} 
\end{center}
\end{figure*}

Figures~\figref{fig9}{(a)}--\figref{fig9}{(c)} show examples of the magnetization dynamics $M^z(t)$ obtained for 12-site spin-$s=1/2$, 6-site spin-$s=1$, and 6-site spin-$s=3/2$ models, respectively. They all require $12$ qubits for the circuit representation. We applied 1DCS by exciting the quantum spin systems with a single magnetic field pulse (shaded area) of strength $B_0=0.5$. The exact simulation results (red circles) obtained by exact diagonalization~\eqref{eq:ED} are presented alongside the corresponding AVQDS results (solid black line). The AVQDS results match well the exact simulation results for all the three models. To assess the accuracy of AVQDS in more detail, the corresponding infidelities $1-f$ are shown in Figs.~\figref{fig9}{(d)}--\figref{fig9}{(f)}. The AVQITE method accurately prepares the ground state with infidelities of $2.9\times 10^{-4}$, $1.1\times 10^{-4}$, and $2.8\times 10^{-5}$ for 12-site spin-$s=1/2$, 6-site spin-$s=1$, and 6-site spin-$s=3/2$ models, respectively. Throughout the time evolution, the infidelity remains below $1.1\times 10^{-3}$ for all three models. Specifically, the infidelity saturates after the pulse excitation for $s=1, 3/2$, while showing a slow increase for $s=1/2$. This demonstrates the capability of the AVQDS approach to accurately simulate the quantum spin dynamics of systems with larger number of sites over long simulation times. 

The required quantum resources in terms of CNOT gates are analyzed in Figs.~\figref{fig9}{(g)}--\figref{fig9}{(i)}. The ground state preparation with the AVQITE method requires 876, 2836, and 5932 CNOT gates for 12-site spin-$s=1/2$, 6-site spin-$s=1$, and 6-site spin-$s=3/2$ models, respectively. Consistent with the two-site spin-model simulations, the number of CNOT gates mainly increases during the $B$-field driving and saturates after a simulation time of about $tJ=15$ for $s=1$ and $tJ=7$ for $s=3/2$, while it shows a slow increase after the pulse for $s=1/2$. The saturated number of CNOTs is 10468, 25066, and 25118 for 12-site spin-$s=1/2$, 6-site spin-$s=1$, and 6-site spin-$s=3/2$ models, respectively.

\begin{figure*}[t!]
\begin{center}
		\includegraphics[scale=0.6]{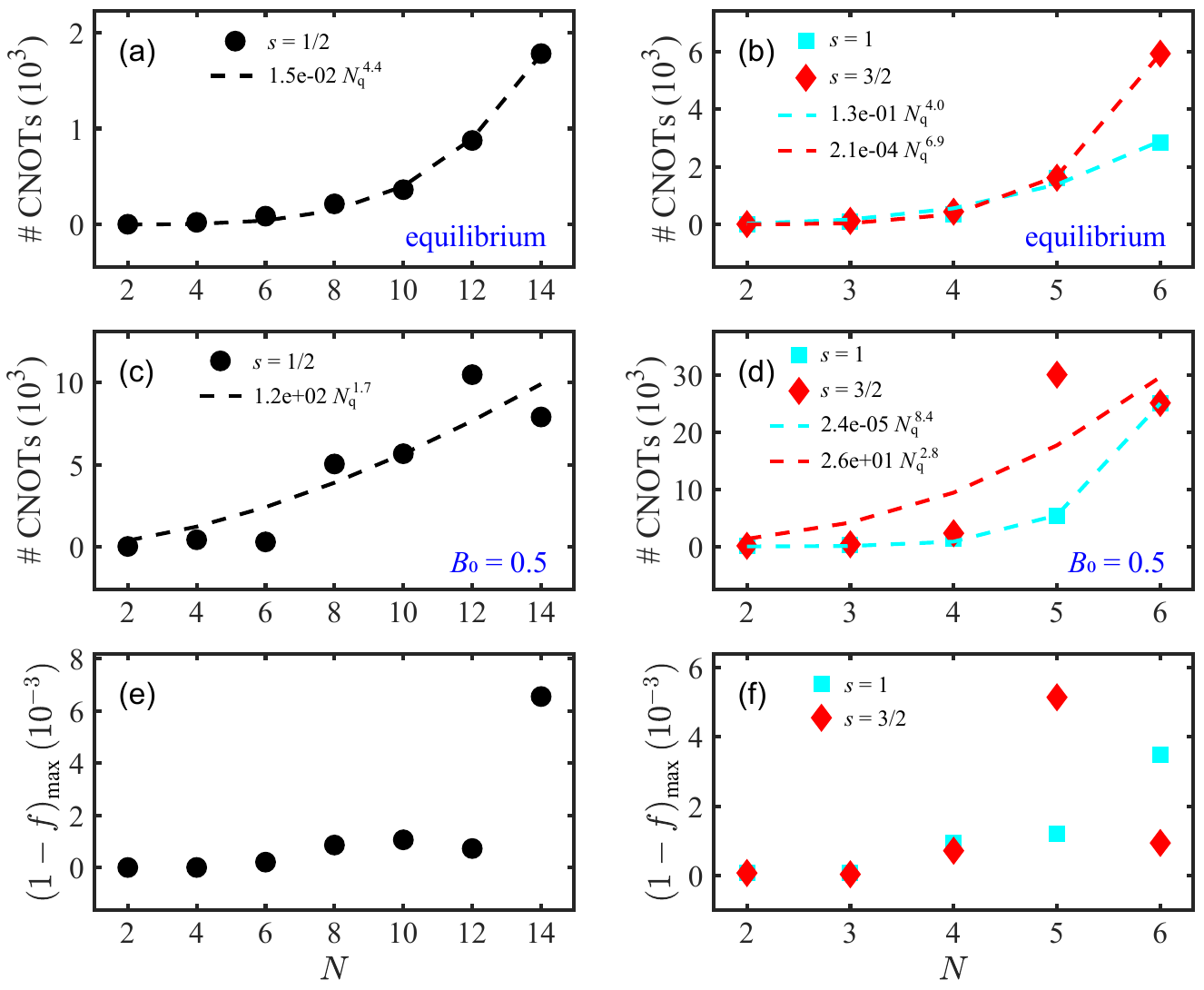}
		\caption{\textbf{Dependence of the initial and saturated number of CNOT gates on the number of sites $N$ for different $s$.} (a), (b) Required number of CNOT gates for the ground state preparation using AVQITE as a function of number of sites $N$ for (a) $s=1/2$ (black circles) and (b) $s=1$ (cyan squares) and $s=3/2$ (red rectangles). Fits using a power-law function of the form $\alpha N^\beta$ are shown for spin $s=1/2$ as dashed black line, for spin $s=1$ as dashed cyan line, and for $s=3/2$ as dashed red line. (c), (d) The corresponding $N$-dependence of the saturated number of CNOT gates for a single magnetic field excitation of strength $B_0=0.5$. (e), (f) The corresponding site-dependence of the maximum infidelity during the time evolution up to a simulation time of $t J=20$.}
		\label{fig10} 
\end{center}
\end{figure*}

The dependence of the required quantum resources on the number of sites $N$ is analyzed in Fig.~\ref{fig10}. The result for ground state preparation using AVQITE is shown in Fig.~\figref{fig10}{(a)} for spin-$s=1/2$ (black circles) and in Fig.~\figref{fig10}{(b)} for spin-$s=1$ (cyan squares) and spin-$s=3/2$ (red rectangles) models. Fitting using a power-law function of the form $\alpha N^\beta$ yields $\beta\approx 4.4$ (dashed black line in Fig.~\figref{fig10}{(a)}), $\beta\approx 4.0$ (dashed cyan line in Fig.~\figref{fig10}{(b)}), and $\beta\approx 6.9$ (dashed red line in Fig.~\figref{fig10}{(b)}) for spin $s=1/2$, $s=1$, and $s=3/2$ models, respectively. Figures~\figref{fig10}{(c)} and \figref{fig10}{(d)} present the corresponding system-size dependence of the saturated number of CNOT gates for a single $B$-field excitation of strength $B_0=0.5$. The fitting with a power-law function results in a scaling of approximately $\propto N_\mathrm{q}^{1.7}$, $\propto N_\mathrm{q}^{8.4}$, and $\propto N_\mathrm{q}^{2.8}$ for spin $s=1/2$, $s=1$, and $s=3/2$ models, respectively. These results imply a polynomial system-size scaling for the quantum resources, although exact scaling cannot be deduced due to limited samples. Finally, we show in Figs.~\figref{fig10}{(e)} and \figref{fig10}{(f)} the site-dependence of the maximum infidelity throughout the time evolution up to a simulation time of $t J=20$, $(1-f)_\mathrm{max}\equiv\max_{t\in [0,20]}[1-f(t)]$, for the three studied spin models. The maximum infidelity increases with growing number of sites but stays below $6.5\times 10^{-3}$ for $s=1/2$, $3.5\times 10^{-3}$ for $s=1$, and $5.1\times 10^{-3}$ for $s=3/2$. This confirms a high accuracy of the AVQDS approach in modelling the magnetic-field driven spin dynamics across a wide range of high-spin systems with system sizes up the $N=14$.

\section{2DCS of two-site spin-$s=5/2$ model \label{sec:4-site}}

\begin{figure*}[t!]
\begin{center}
		\includegraphics[scale=0.57]{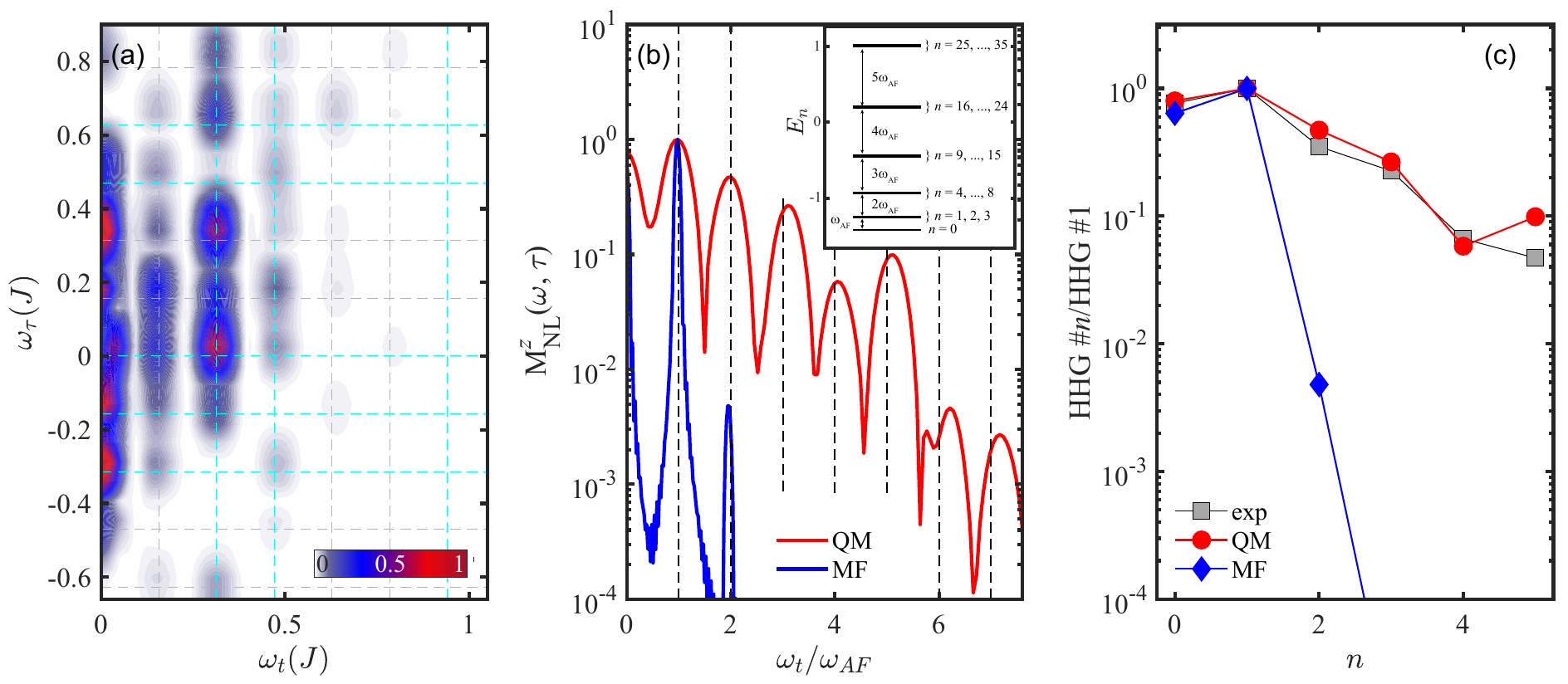}
		\caption{\textbf{Two-dimensional coherent spectroscopy of two-site spin-$s=5/2$ system.}  (a) 2DCS spectrum for $B_0=4.0$ shows distinct peaks at multiples of the magnon frequency $\omega_\mathrm{AF}$ along both $\omega_t$- and $\omega_\tau$-directions indicated by dashed lines. (b) $M^z_\mathrm{NL}(\omega_t,\tau)$ spectrum at fixed inter-pulse delay of $\tau J=7.6$. The result of the full quantum spin simulation (red line) is compared with the corresponding result of the mean-field theory (blue line). The quantum spin simulation shows high-harmonic generation peaks (vertical dashed lines) up to seventh-harmonic generation while the signals are weaker for the mean-field simulation. Inset: Eigenenergies $E_n$ of the unperturbed Hamiltonian $\hat{\mathcal{H}}_0$. The two-site spin-$s=5/2$ has $(2s+1)^2=36$ eigenstates. Differences between the eigenenergies approximately correspond to multiples of the magnon frequency $\omega_\mathrm{AF}\approx 0.16$ (c) Ratio between the strength of $n$th harmonic and fundamental harmonic generation peaks. The high-harmonic generation strengths deduced from the experiment (squares) are shown together with the result of the quantum spin simulation (circles) and the mean-field simulation (rectangles). The strength of the magnonic high-harmonic generation signals in the simulated 2DCS spectra of the quantum high-spin model agrees well with the experiment, in contrast to the corresponding result of the mean-field model.}
		\label{fig8} 
\end{center}
\end{figure*}

In this section, we study 2DCS of a two-site spin-$s=5/2$ model and compare with the results of a 2DCS experiment performed on a rare-earth orthoferrite material~\cite{Zhao2016}. In the simulations we use an antiferromagnetic exchange interaction of $J_1=1$ in the Hamiltonian~(\ref{eq:Ham}). We set $K_a=0.0012$ and $K_c=0.0006$ according to the values derived from inelastic neutron scattering measurements on rare-earth orthoferrite materials~\cite{Hahn:2014}; while we adjust the DM interaction strength to $D=0.2$ to match the strong nonlinearities observed in the 2DCS experiment, which is about one order of magnitude larger than the derived value from neutron scattering.
In Fig.\figref{fig8}{(a)} we show the 2DCS spectrum for a magnetic field strength of $B_0=4.0$, which is about $4$-times larger than the strength used in the experiment. The spectrum shows peaks at multiples of the magnon frequency $\omega_\mathrm{AF}$ along both $\omega_t$- and $\omega_\tau$-directions indicated by dashed lines which result from transitions between different eigenstates based on the discussion in section~\ref{sec:1Dvs2D}. Here we note that even though the spectrum of the eigenenergies in the inset of Fig.\figref{fig8}{(b)} is not harmonic, the difference between eigenenergies corresponds to multiples of the magnon frequency $\omega_\mathrm{AF}$. The 2DCS spectrum in Fig.\figref{fig8}{(a)} shows high-harmonic generation up to fourth order along $\omega_t$, e.~g., at fixed $\omega_\tau=\omega_\mathrm{AF}$.  To study the magnonic high-harmonic generation in more detail, we plot in Fig.\figref{fig8}{(b)} the $M^z_\mathrm{NL}(\omega_t,\tau)$ spectrum at fixed inter-pulse delay of $\tau J = 7.6$ which approximately corresponds to in-phase magnon excitation by the magnetic field pulse pair. This spectrum shows high-harmonic generation peaks (vertical dashed lines) up to seventh-harmonic generation centered at frequencies of up to seven times the magnon frequency.  In Fig.\figref{fig8}{(b)} we also show the result of the corresponding mean-field simulation, which is discussed in more detail in appendix~\ref{sec:MF}. The high harmonic generation peaks are much weaker compared to the result of the full quantum spin model.

Finally, we compare the simulation results with THz 2DCS measurement performed on a rare-earth orthoferrite $a$-cut \(\text{Sm}_{0.4}\text{Er}_{0.6}\text{FeO}_{3}\) sample at room temperature, where the system shows canted antiferromagnetic order with a small net magnetization along $c$-axis~\cite{Zhao2016}. The sample is excited by two collinear THz magnetic pulses polarized parallel to sample $c$-axis. The transmitted nonlinear emission is recorded as a function of gate time at fixed inter-pulse delay $\tau$ using electric-optical sampling. To compare the simulation results with the experiment, we show in Fig.\figref{fig8}{(c)} the ratio between the peak strengths of the $n$th harmonic and the fundamental harmonic. The ratios extracted from the experiment (squares) are shown together with the results of the quantum spin (circles) and mean-field (rectangles) simulations. The strengths of the high-harmonic generation peaks of the quantum-spin model are in good agreement with the results of the experiment. Compared to that, the result based on the mean-field description of the spins discussed in appendix~\ref{sec:MF} yields much smaller high-harmonic generation signals. This implies that a quantum spin modelling is required to explain the strong magnonic high-harmonic generation signals in the experiment. It is important to note that the simple two-site quantum spin model studied in this section does not offer quantitative agreement with the experiment across arbitrary time delays. However, all time delays indicate enhancement of high-harmonic generation peak strengths compared to the mean-field model, albeit with a lesser degree of agreement than the time trace studied in Fig.~\ref{fig8}. For a more quantitative comparison of the entire two-dimensional spectrum with the experiment, quantum spin calculations with $N=4$-sites and beyond in the Hamiltonian~\eqref{eq:Ham} are required.

\section{Conclusion and outlook}
In this work, we utilize the AVQDS approach to investigate two-dimensional coherent spectra of a high spin-model. The model we consider is an antiferromagnetic two-site quantum high-spin Hamiltonian, which incorporates Dzyaloshinskii-Moriya interactions and single-ion anisotropies. To transform the high-spin operators of the Hamiltonian into multi-qubit operators, we apply two encodings: the standard binary encoding and Gray code. We first calculate the 1DCS and 2DCS spectra for a reduced two-site spin-$s=1$ model. By comparing the AVQDS results with exact simulation results, we demonstrate that the AVQDS simulation accurately captures the quantum spin dynamics, with adaptively generated quantum circuits with less than $100$ CNOT gates. The obtained 2DCS spectra reveal peaks at multiples of the magnon frequency in both the $\omega_t$ and $\omega_\tau$ directions. By employing a susceptibility expansion of the nonlinear magnetization, we illustrate that these peaks in the 2D frequency space arise from transitions between different eigenstates of the unperturbed Hamiltonian. Furthermore, we show that 2DCS offers a higher resolution of the energy spectrum of the high spin model compared to 1DCS.

We then proceeded to investigate how the required quantum resources change with spin $s$ for the standard binary encoding and Gray code. Both encodings necessitate a comparable number of CNOT gates to prepare the initial ground state using the qubit-ADAPT-VQE approach, exhibiting polynomial scaling of $N_\mathrm{q}^{4.3}$ for the standard binary encoding and $N_\mathrm{q}^{4.5}$ for the Gray code. The scaling of the saturated number of CNOT gates for low magnetic field driving is similar for both encodings, displaying polynomial scaling of $N_\mathrm{q}^{3.8}$ and $N_\mathrm{q}^{3.7}$ for the standard binary encoding and Gray code, respectively. In contrast, the Gray code requires fewer quantum resources for strong magnetic field driving, with the saturated number of CNOT gates scaling as $N_\mathrm{q}^{4.5}$ compared to $N_\mathrm{q}^{5.0}$ for the standard binary encoding. The obtained polynomial scaling is comparable to the scaling obtained for the nonintegrable mixed-field Ising model in~\cite{AVQDS} where the number of CNOT gates scales as $N_\mathrm{q}^5$  which demonstrates the complexity of the studied high-spin model. We further performed simulations for quantum spin models with system-size up to $N=14$ for spin $s=1/2$ and $N=6$ for spins $s=1, 3/2$. The results indicate a polynomial system-size scaling for the quantum resources, although exact scaling cannot be deduced due to limited samples.

Finally, we calculated the 2DCS spectrum for a two-site spin-$s=5/2$ model to compare it with the results of an experiment conducted on a rare-earth orthoferrite. The strengths of the observed magnonic high-harmonic generation signals in the 2DCS spectra of the quantum high-spin model are in good agreement with the experimental results, in contrast to the corresponding mean-field model. This finding suggests that a complete quantum spin model is necessary to explain the strong magnonic high-harmonic generation signals observed in the experiment.

As a further direction, the required number of CNOT gates for the AVQDS approach to simulate quantum spin systems of larger sizes is still relatively large, demanding further algorithmic development to reduce quantum resources. Here, alternative qudit-based architectures~\cite{wang2020, Ogunkoya2024QutritCA} or bosonic quantum processors~\cite{dutta2024simulating, stavenger2022c2qa}, rather than qubit-based platform, could be more advantageous in simulating higher spin models, since the overhead of Hamiltonian encoding is removed by the $d$-level qudit with $d=2s+1$ for spin magnitude $s$. In addition, the AVQDS approach could be further improved by incorporating multiple disjoint generators in each circuit layer, similar to the tetris-ADAPT VQE approach~\cite{anastasiou2022}, which might further reduce the circuit depth for high-spin model simulations. In order to deploy the AVQDS approach in practical calculations on quantum hardware, it is crucial to consider the implications of hardware noise and statistical errors arising from finite number of measurements. Additionally, incorporating error mitigation techniques becomes necessary in this context~\cite{Viola1998DYNAMICALSO, ZNE_Temme, Li_Benjamin-PRX-2017,caiQuantumErrorMitigation2022,vandenbergProbabilisticErrorCancellation2023,McDonough-AutomatedPER-2022}.



\section*{Acknowledgements}
Y.Y acknowledges useful discussions with M. DeMarco, K. Smith, E. Crane, and T.-C. Wei. This project was primarily supported by the U.S. Department of Energy, Office of Science, National Quantum Information Science Research Centers, Co-design Center for Quantum Advantage under contract number DE-SC0012704. The algorithmic and code development of AVQDS was supported by the U.S. Department of Energy (DOE), Office of Science, Basic Energy Sciences, Materials Science and Engineering Division, including the grant of computer time at the National Energy Research Scientific Computing Center (NERSC) in Berkeley, California. The research was performed at the Ames National Laboratory, which is operated for the U.S. Department of Energy by Iowa State University under Contract No.~DE-AC02-07CH11358. The 2DCS experiment for the Sm$_{0.4}$Er$_{0.6}$FeO$_{3}$ sample performed by C.H., L.L., and J.W. was supported by the U.S. Department of Energy (DOE), Office of Science, Basic Energy Sciences, Materials Science and Engineering Division. P.P.O. was supported by the U.S. Department of Energy, Office of Science, National Quantum Information Science Research Centers, Superconducting Quantum Materials and Systems Center (SQMS) under the contract No. DE-AC02- 07CH11359. P.P.O. acknowledges useful discussions with Y.~Qiang, V.~L.~Quito, and T.~V.~Trevisan.



\appendix

\section{Binary encodings of high-spin operators \label{sec:S_trafo}}

In this appendix we present the explicit transformations of high-spin $s$ operators to multi-qubit operators based on section~\ref{sec:S_trans} for $s=1,\,3/2,\,2$, and $5/2$. Using the standard binary enconding, the spin-$s=1$ operators can be written as
\begin{align}
	&\hat{S}^x=\frac{1}{\sqrt{8}}\left[\sigma_x^{(0)}+\sigma_x^{(1)}\sigma_x^{(0)}+\sigma_y^{(1)}\sigma_y^{(0)}+\sigma_z^{(1)}\sigma_x^{(0)}\right]\,,\nonumber \\	&\hat{S}^y=\frac{1}{\sqrt{8}}\left[\sigma_y^{(0)}-\sigma_x^{(1)}\sigma_y^{(0)}+\sigma_y^{(1)}\sigma_x^{(0)}+\sigma_z^{(1)}\sigma_y^{(0)}\right]\,, \nonumber \\
	&\hat{S}^z=\frac{1}{2}\left[\sigma_z^{(1)}+\sigma_z^{(1)}\sigma_z^{(0)}\right]\,.	
\end{align}
For $s=3/2$ we find
\begin{align}
    &\hat{S}^x=\frac{1}{2}\left[\sqrt{3}\,\sigma_x^{(0)}+\sigma_x^{(1)}\sigma_x^{(0)}+\sigma_y^{(1)}\sigma_y^{(0)}\right]\,,\nonumber \\
    &\hat{S}^y=\frac{1}{2}\left[\sqrt{3}\,\sigma_y^{(0)}-\sigma_x^{(1)}\sigma_y^{(0)}+\sigma_y^{(1)}\sigma_x^{(0)}\right]\,,\nonumber \\   
    &\hat{S}^z=\frac{1}{2}\sigma_z^{(0)}+\sigma_z^{(1)}\,.
\end{align}
The spin-$s=2$ operators become
\begin{widetext}
\begin{align}
     &\hat{S}^x=\frac{1}{4}\left(1+\sqrt{\frac{3}{2}}\right)\sigma_x^{(0)}+\frac{1}{4}\sqrt{\frac{3}{2}}\left(\sigma_x^{(1)}\sigma_x^{(0)}+\sigma_y^{(1)}\sigma_y^{(0)}-\sigma_z^{(1)}\sigma_x^{(0)}+\sigma^{(2)}_z\sigma_x^{(0)}\right)+\frac{1}{4}\left(\sigma_z^{(1)}\sigma_x^{(0)}+\sigma_z^{(2)}\sigma_x^{(0)}\right)\nonumber \\
	&\qquad\quad +\frac{1}{4}\left(\sigma_x^{(2)}\sigma_x^{(1)}\sigma_x^{(0)}-\sigma_x^{(2)}\sigma_y^{(1)}\sigma_y^{(0)}+\sigma_y^{(2)}\sigma_x^{(1)}\sigma_y^{(0)}+\sigma_y^{(2)}\sigma_y^{(1)}\sigma_x^{(0)}+\sigma_z^{(2)}\sigma_z^{(1)}\sigma_x^{(0)}\right)\nonumber \\
	&\qquad\quad +\frac{1}{4}\sqrt{\frac{3}{2}}\left(\sigma_z^{(2)}\sigma_x^{(1)}\sigma_x^{(0)}+\sigma_z^{(2)}\sigma_y^{(1)}\sigma_y^{(0)}-\sigma_z^{(2)}\sigma_z^{(1)}\sigma_x^{(0)}\right)\,,\nonumber \\
    &\hat{S}^y=\frac{1}{4}\left(1+\sqrt{\frac{3}{2}}\right)\sigma_y^{(0)}+\frac{1}{4}\sqrt{\frac{3}{2}}\left(\sigma_y^{(1)}\sigma_x^{(0)}-\sigma_x^{(1)}\sigma_y^{(0)}-\sigma_z^{(1)}\sigma_y^{(0)}+\sigma^{(2)}_z\sigma_y^{(0)}\right)+\frac{1}{4}\left(\sigma_z^{(1)}\sigma_y^{(0)}+\sigma_z^{(2)}\sigma_y^{(0)}\right)\nonumber \\
	&\qquad\quad +\frac{1}{4}\left(\sigma_y^{(2)}\sigma_x^{(1)}\sigma_x^{(0)}-\sigma_x^{(2)}\sigma_x^{(1)}\sigma_y^{(0)}-\sigma_x^{(2)}\sigma_y^{(1)}\sigma_x^{(0)}-\sigma_y^{(2)}\sigma_y^{(1)}\sigma_y^{(0)}+\sigma_z^{(2)}\sigma_z^{(1)}\sigma_y^{(0)}\right)\nonumber \\
	&\qquad\quad +\frac{1}{4}\sqrt{\frac{3}{2}}\left(\sigma_z^{(2)}\sigma_y^{(1)}\sigma_x^{(0)}-\sigma_z^{(2)}\sigma_x^{(1)}\sigma_y^{(0)}-\sigma_z^{(2)}\sigma_z^{(1)}\sigma_y^{(0)}\right)\,,\nonumber \\  
    &\hat{S}^z=\frac{1}{4}\sigma_z^{(1)}+\frac{1}{2}\sigma_z^{(2)}-\frac{1}{4}\sigma_z^{(1)}\sigma_z^{(0)}+\frac{1}{2}\sigma_z^{(2)}\sigma_z^{(0)}+\frac{3}{4}\sigma_z^{(2)}\sigma_z^{(1)}+\frac{1}{4}\sigma_z^{(2)}\sigma_z^{(1)}\sigma_z^{(0)}\,,   
\end{align}
\end{widetext}
while the spin-$s=5/2$ operators take the form
\begin{widetext}
\begin{align}
	&\hat{S}^x=\frac{3+2\sqrt{5}}{8}\sigma_x^{(0)}+\frac{1}{\sqrt{8}}\left(\sigma_x^{(1)}\sigma_x^{(0)}+\sigma_y^{(1)}\sigma_y^{(0)}\right)+\frac{3}{8}\left(\sigma_z^{(2)}\sigma_x^{(0)}-\sigma_z^{(1)}\sigma_x^{(0)}\right)+\frac{\sqrt{5}}{4}\sigma_z^{(1)}\sigma_x^{(0)}\nonumber \\
	&\qquad\quad +\frac{1}{\sqrt{8}}\left(\sigma_x^{(2)}\sigma_x^{(1)}\sigma_x^{(0)}-\sigma_x^{(2)}\sigma_y^{(1)}\sigma_y^{(0)}+\sigma_y^{(2)}\sigma_x^{(1)}\sigma_y^{(0)}+\sigma_y^{(2)}\sigma_y^{(1)}\sigma_x^{(0)}+\sigma_z^{(2)}\sigma_x^{(1)}\sigma_x^{(0)}+\sigma_z^{(2)}\sigma_y^{(1)}\sigma_y^{(0)}\right)-\frac{3}{8}\sigma_z^{(2)}\sigma_z^{(1)}\sigma_x^{(0)}\,, \nonumber \\
	&\hat{S}^y=\frac{3+2\sqrt{5}}{8}\sigma_y^{(0)}+\frac{1}{\sqrt{8}}\left(\sigma_y^{(1)}\sigma_x^{(0)}-\sigma_x^{(1)}\sigma_y^{(0)}\right)+\frac{3}{8}\left(\sigma_z^{(2)}\sigma_y^{(0)}-\sigma_z^{(1)}\sigma_y^{(0)}\right)+\frac{\sqrt{5}}{4}\sigma_z^{(1)}\sigma_y^{(0)}\nonumber \\
	&\qquad\quad +\frac{1}{\sqrt{8}}\left(-\sigma_y^{(2)}\sigma_y^{(1)}\sigma_y^{(0)}+\sigma_y^{(2)}\sigma_x^{(1)}\sigma_x^{(0)}-\sigma_x^{(2)}\sigma_y^{(1)}\sigma_x^{(0)}-\sigma_x^{(2)}\sigma_x^{(1)}\sigma_y^{(0)}+\sigma_z^{(2)}\sigma_y^{(1)}\sigma_x^{(0)}-\sigma_z^{(2)}\sigma_x^{(1)}\sigma_y^{(0)}\right)-\frac{3}{8}\sigma_z^{(2)}\sigma_z^{(1)}\sigma_y^{(0)}\,,\nonumber \\
	&\hat{S}^z=\frac{3}{8}\sigma_z^{(0)}+\sigma_z^{(2)}+\frac{1}{8}\left(\sigma_z^{(2)}\sigma_z^{(0)}+\sigma_z^{(1)}\sigma_z^{(0)}\right)+\sigma_z^{(2)}\sigma_z^{(1)}-\frac{1}{8}\sigma_z^{(2)}\sigma_z^{(1)}\sigma_z^{(0)}\,.
\end{align}
\end{widetext}

Using the Gray code, the spin-$s=1$ operators are encoded via
\begin{align}
	&\hat{S}^x=\frac{1}{\sqrt{8}}\left[\sigma_x^{(0)}+\sigma_x^{(1)}-\sigma_x^{(1)}\sigma_z^{(0)}+\sigma_z^{(1)}\sigma_x^{(0)}\right]\,,\nonumber \\
	&\hat{S}^y=\frac{1}{\sqrt{8}}\left[\sigma_y^{(0)}+\sigma_y^{(1)}-\sigma_y^{(1)}\sigma_z^{(0)}+\sigma_z^{(1)}\sigma_y^{(0)}\right]\,, \nonumber \\
	&\hat{S}^z=\frac{1}{2}\left[\sigma_z^{(0)}+\sigma_z^{(1)}\right]\,.	
\end{align}
For $s=3/2$ the spin operators read
\begin{align}
    &\hat{S}^x=\frac{1}{2}\left[\sqrt{3}\,\sigma_x^{(0)}+\sigma_x^{(1)}-\sigma_x^{(1)}\sigma_z^{(0)}\right]\,,\nonumber \\
    &\hat{S}^y=\frac{1}{2}\left[\sigma_y^{(1)}-\sigma_y^{(1)}\sigma_z^{(0)}+\sqrt{3}\,\sigma_z^{(1)}\sigma_y^{(0)}\right]\,,\nonumber \\   
    &\hat{S}^z=\sigma_z^{(1)}+\frac{1}{2}\sigma_z^{(1)}\sigma_z^{(0)}\,.
\end{align}
The spin-$s=2$ operators take the form
\begin{widetext}
\begin{align}
     &\hat{S}^x=\frac{1}{4}\left(1+\sqrt{\frac{3}{2}}\right)\sigma_x^{(0)}+\frac{1}{4}\sqrt{\frac{3}{2}}\sigma_x^{(1)}+\frac{1}{4}\sigma_x^{(2)}+\frac{1}{4}\sqrt{\frac{3}{2}}\left(\sigma_z^{(2)}\sigma_x^{(0)}+\sigma_z^{(2)}\sigma_x^{(1)}-\sigma_x^{(1)}\sigma_z^{(0)}-\sigma_z^{(1)}\sigma_x^{(0)}\right)\nonumber \\
	&\qquad\quad +\frac{1}{4}\left(\sigma_x^{(2)}\sigma_z^{(0)}+\sigma_z^{(1)}\sigma_x^{(0)}-\sigma_x^{(2)}\sigma_z^{(1)}+\sigma_z^{(2)}\sigma_x^{(0)}\right)-\frac{1}{4}\sqrt{\frac{3}{2}}\left(\sigma_z^{(2)}\sigma_x^{(1)}\sigma_z^{(0)}+\sigma_z^{(2)}\sigma_z^{(1)}\sigma_x^{(0)}\right)\nonumber \\
	&\qquad\quad+\frac{1}{4}\left(\sigma_z^{(2)}\sigma_z^{(1)}\sigma_x^{(0)}-\sigma_x^{(2)}\sigma_z^{(1)}\sigma_z^{(0)}\right)\,,\nonumber \\ 
     &\hat{S}^y=\frac{1}{4}\left(1-\sqrt{\frac{3}{2}}\right)\sigma_y^{(0)}+\frac{1}{4}\sqrt{\frac{3}{2}}\sigma_y^{(1)}+\frac{1}{4}\sigma_y^{(2)}+\frac{1}{4}\sqrt{\frac{3}{2}}\left(\sigma_z^{(2)}\sigma_y^{(0)}-\sigma_y^{(1)}\sigma_z^{(0)}-\sigma_z^{(2)}\sigma_y^{(0)}+\sigma_z^{(2)}\sigma_y^{(1)}\right)\nonumber \\
	&\qquad\quad +\frac{1}{4}\left(\sigma_y^{(2)}\sigma_z^{(0)}+\sigma_z^{(1)}\sigma_y^{(0)}-\sigma_y^{(2)}\sigma_z^{(1)}+\sigma_z^{(2)}\sigma_y^{(0)}\right)+\frac{1}{4}\sqrt{\frac{3}{2}}\left(\sigma_z^{(2)}\sigma_z^{(1)}\sigma_y^{(0)}-\sigma_z^{(2)}\sigma_y^{(1)}\sigma_z^{(0)}\right)\nonumber \\
	&\qquad\quad+\frac{1}{4}\left(\sigma_z^{(2)}\sigma_z^{(1)}\sigma_y^{(0)}-\sigma_y^{(2)}\sigma_z^{(1)}\sigma_z^{(0)}\right)\,,\nonumber \\  
    &\hat{S}^z=-\frac{1}{4}\sigma_z^{(0)}+\frac{3}{4}\sigma_z^{(1)}+\frac{1}{2}\sigma_z^{(2)}+\frac{1}{2}\sigma_z^{(1)}\sigma_z^{(0)}+\frac{1}{4}\sigma_z^{(2)}\sigma_z^{(0)}+\frac{1}{4}\sigma_z^{(2)}\sigma_z^{(1)}\,,   
\end{align}
\end{widetext}
while the spin-$s=5/2$ operators become
\begin{widetext}
\begin{align}
	&\hat{S}^x=\frac{1}{8}\left[\left(2\sqrt{5}+3\right)\sigma_x^{(0)}+\sqrt{8}\left(\sigma_x^{(1)}+\sigma_x^{(2)}\right)+\sqrt{8}\left(\sigma_z^{(2)}\sigma_x^{(1)}-\sigma_x^{(1)}\sigma_z^{(0)}-\sigma_x^{(2)}\sigma_z^{(1)}+\sigma_x^{(2)}\sigma_z^{(0)}\right)\right.\nonumber \\
	&\qquad\quad\left. +3\left(\sigma_z^{(2)}\sigma_x^{(0)}-\sigma_z^{(1)}\sigma_x^{(0)}\right)+\left(2\sqrt{5}-3\right)\sigma_z^{(2)}\sigma_z^{(1)}\sigma_x^{(0)}-\sqrt{8}\left(\sigma_z^{(2)}\sigma_x^{(1)}\sigma_z^{(0)}+\sigma_x^{(2)}\sigma_z^{(1)}\sigma_z^{(0)}\right)\right]\,,\nonumber \\
	&\hat{S}^y=\frac{1}{8}\left[\left(2\sqrt{5}-3\right)\sigma_y^{(0)}+\sqrt{8}\left(\sigma_y^{(1)}+\sigma_y^{(2)}\right)+\sqrt{8}\left(\sigma_z^{(2)}\sigma_y^{(1)}-\sigma_y^{(1)}\sigma_z^{(0)}-\sigma_y^{(2)}\sigma_z^{(1)}+\sigma_y^{(2)}\sigma_z^{(0)}\right)\right.\nonumber \\
	&\qquad\quad\left. -3\left(\sigma_z^{(2)}\sigma_y^{(0)}-\sigma_z^{(1)}\sigma_y^{(0)}\right)+\left(2\sqrt{5}+3\right)\sigma_z^{(2)}\sigma_z^{(1)}\sigma_y^{(0)}-\sqrt{8}\left(\sigma_z^{(2)}\sigma_y^{(1)}\sigma_z^{(0)}+\sigma_y^{(2)}\sigma_z^{(1)}\sigma_z^{(0)}\right)\right]\,, \nonumber \\
	&\hat{S}^z=\frac{1}{8}\sigma_z^{(0)}+\sigma_z^{(1)}+\sigma_z^{(2)}+\frac{1}{8}\left(\sigma_z^{(1)}\sigma_z^{(0)}-\sigma_z^{(2)}\sigma_z^{(0)}\right)+\frac{3}{8}\sigma_z^{(2)}\sigma_z^{(1)}\sigma_z^{(0)}\,.	
\end{align}
\end{widetext}

\section{AVQDS with Euler vs. fourth-order Runge-Kutta integrator \label{sec:Euler}}

\begin{figure}[t!]
\begin{center}
		\includegraphics[scale=0.8]{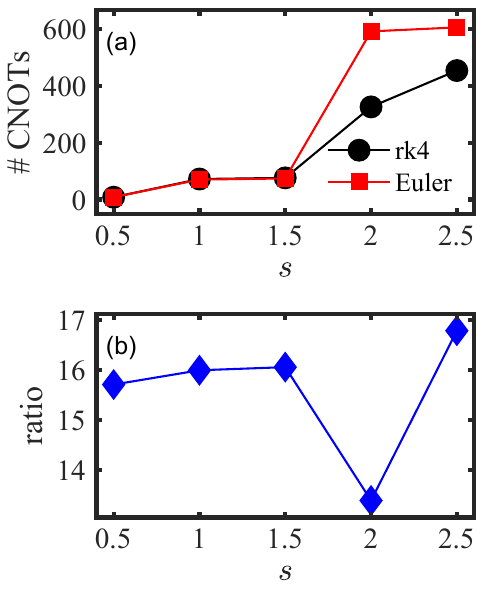}
		\caption{\textbf{Performance of AVQDS with Euler vs. fourth-order Runge-Kutta integrator}. (a)  Saturated number of CNOT gates as a function of spin $s$ for AVQDS where the variational parameter dynamics is calculated using the fourth-order Runge-Kutta (rk4, black circles) and  Euler method (red squares). AVQDS with Runge-Kutta method requires less CNOT gates for high spin $s$. (b) Ratio between total number of time steps of the Euler and Runge-Kutta method. The Euler method needs 13-to-17 times more time steps compared to the Runge-Kutta method to obtain accurate results.}
		\label{fig1_app} 
\end{center}
\end{figure}

In this appendix we compare the performance of AVQDS when integrating the equation of motion~(\ref{eq:ode}) using the Euler method versus a fourth-order Runge-Kutta method. Figure\figref{fig1_app}{(a)} presents the $s$-dependence of the saturated number of CNOT gates for AVQDS where the variational parameter dynamics is calculated using the fourth-order Runge-Kutta (rk4, black circles) and Euler method (red squares). While the number of needed quantum resources is comparable for spin numbers up to $s=3/2$, the fourth-order Runge-Kutta method requires significantly lower number of CNOT gates for $s=2$ and $s=5/2$. The latter results from the more accurate simulation of the long-time quantum spin dynamics with the fourth-order Runge-Kutta integrator compared to the Euler method. As a result, the Runge-Kutta method yields shallower quantum circuits for $s \geq 2$. In addition, the Euler method requires choosing a smaller step size $\delta t$ to obtain accurate results. As discussed in section~\ref{sec:AVQDS}, $\delta t$ is dynamically adjusted and set by a predefined maximum step size $\delta\theta_\mathrm{max}$. We find that $\delta\theta_\mathrm{max}=5.0\times 10^{-3}$ generates accurate results for the Runge-Kutta method, while one needs to use a smaller value $\delta\theta_\mathrm{max}=2.5\times 10^{-4}$ for the  Euler method to reach similar accuracy. To quantify the difference in the step size between both methods in more detail, we calculate the total number of time steps, $N_\mathrm{t}$, during the complete time evolution for both methods. Figure\figref{fig1_app}{(b)} shows the ratio between $N_\mathrm{t}$ of the Euler and Runge-Kutta method. The Euler method needs 13-to-17 times more time steps compared to the Runge-Kutta method to produce accurate results. As a result, even though the fourth-order Runge-Kutta method is computationally more expensive than Euler's method, it is quantum computationally more efficient.

\section{Mean-field description \label{sec:MF}}

Mean-field decoupling of the two-site Hamiltonian~\eqref{eq:Ham} leads to
\begin{align}
    \hat{\mathcal{H}}_\mathrm{MF}&=J_1(\mathbf{m}_0\cdot\hat{\mathbf{S}}_1+\hat{\mathbf{S}}_0\cdot\mathbf{m}_1)\nonumber \\
    &-D(m^z_0\hat{S}^x_1+\hat{S}^z_0 m^x_1-m_0^x\hat{S}^z_1-\hat{S}_0^x m^z_1)\nonumber \\
    &-2\sum_{i=0}^{N-1}\left[K_a m^x_i \hat{S}^x_i+K_c m^z_i \hat{S}^z_i\right]-B\sum_{i=0}^{N-1}\hat{S}^z_i\,,
    \label{eq:Ham_MF}
\end{align}
where $\mathbf{m}_i=\langle \hat{\mathbf{S}}_i\rangle$. To obtain the ground state of the mean-field Hamiltonian, we make the ansatz 
\begin{align}
    \mathbf{m}_i=s((-1)^i\sin\varphi,0,\cos\varphi)\,.
    \label{eq:ansatz2}
\end{align}
Expressing the classical energy $E=\langle \hat{\mathcal{H}}_\mathrm{MF}\rangle$ for $B=0$ in terms of the ansatz~\eqref{eq:ansatz2} yields
\begin{align}
    E&=2J_1s^2\cos 2\varphi+2Ds^2\sin 2\varphi \nonumber \\
    &-4K_a s^2 \sin^2\varphi-4K_c s^2\cos^2\varphi\,.
\end{align}
Minimization of $E$ with respect to $\varphi$ produces
\begin{align}
    \tan 2\varphi=\frac{D}{J_1+K_a-K_c}\,,
\end{align}
which determines the spin canting angle of the ground state spin configuration of the mean-field Hamiltonian. To obtain the dynamics of $M^z(t)$ for the mean-field Hamiltonian~\eqref{eq:Ham_MF}, we self-consistently solve the time-dependent Schr\"odinger equation $\partial_t|\Psi[t]\rangle=-i\,\hat{\mathcal{H}}_\mathrm{MF}(t)|\Psi[t]\rangle$ and $\mathbf{m}_j(t)=\langle\Psi[t]|\hat{\mathbf{S}}_j|\Psi[t]\rangle$ using a fourth-order Runge-Kutta method.


\bibliography{ref}

\end{document}